\documentclass[fleqn,usenatbib]{mnras}

\usepackage{newtxtext,newtxmath}

\usepackage[T1]{fontenc}
\usepackage{ae,aecompl}

\usepackage{graphicx}	
\usepackage{amsmath}	
\usepackage{siunitx}    

\title[On the density regime probed by HCN emission]{On the density regime probed by HCN emission}

\author[G. H. Jones et al.]{
Gerwyn H. Jones,$^{1}$\thanks{E-mail: Gerwyn.Jones@astro.cf.ac.uk (GHJ)}
Paul C. Clark,$^{1}$\thanks{E-mail: ClarkPC@cardiff.ac.uk (PCC)}
Simon C. O. Glover$^{2}$,
Alvaro Hacar$^{3}$
\\
$^{1}$School of Physics and Astronomy, Queen's Buildings, The Parade, Cardiff University, Cardiff, CF24 3AA\\
$^{2}$Zentrum f{\"u}r Astronomie der Universit{\"a}t Heidelberg, Institut f{\"u}r Theoretische Astrophysik, Albert-Ueberle-Str. 2, D-69120 Heidelberg, Germany\\
$^{3}$University of Vienna, Department of Astrophysics, T{\"u}rkenschanzstrasse 17, A-1180 Vienna, Austria
}

\date{Accepted XXX. Received YYY; in original form ZZZ}

\pubyear{2021}

\begin{document}
\label{firstpage}
\pagerange{\pageref{firstpage}--\pageref{lastpage}}
\maketitle

\begin{abstract}
HCN J$\, =\,$1$\, -\,$0 emission is commonly used as a dense gas tracer, thought to mainly arise from gas with densities $\mathrm{\sim 10^4\ -\ 10^5\ cm^{-3}}$. This has made it a popular tracer in star formation studies. However, there is increasing evidence from observational surveys of `resolved' molecular clouds that HCN can trace more diffuse gas. We investigate the relationship between gas density and HCN emission through post-processing of high resolution magnetohydrodynamical simulations of cloud-cloud collisions. We find that HCN emission traces gas with a mean volumetric density of $\mathrm{\sim 3 \times 10^3\ cm^{-3}}$ and a median visual extinction of $\mathrm{\sim 5\ mag}$. We therefore predict a characteristic density that is an order of magnitude less than the ``standard'' characteristic density of $\mathrm{n \sim 3 \times 10^4\ cm^{-3}}$. Indeed, we find in some cases that there is clear HCN emission from the cloud even though there is no gas denser than this standard critical density.   We derive independent luminosity-to-mass conversion factors for the amount of gas at $\mathrm{A_V} > 8$ or at densities $\mathrm{n > 2.85 \times 10^{3} \: cm^{-3}}$ or $\mathrm{n > 3 \times 10^{4} \: cm^{-3}}$, finding values of $\alpha_{\rm HCN} = 6.79, 8.62$ and $27.98 \: {\rm M_{\odot}} ({\rm K \, km \, s^{-1} \, pc^{2}})$, respectively. 

\end{abstract}




\section{Introduction}
\label{sec:intro}

A central goal of star formation theory is to predict the rate at which gas in the interstellar medium (ISM) is converted into stars,  and so research into this field has focused on measuring the amount of gas present to form stars, and connecting that to the amount of star formation that actually occurs.  
Perhaps the most widely-studied relation is that between the surface density of the star formation rate, and the surface density of gas, known as the Kennicutt - Schmidt (K-S) relation \citep{SCHMIDT1959, KENNICUTT1989},
\begin{equation}
    \mathrm{\Sigma_{SFR} \propto \Sigma_{gas}^N }.
\end{equation}
 \cite{KENNICUTT1989} found N to be $\mathrm{1.4 \pm 0.15}$, however there has been some debate over the value of the index since \citep{BIGIEL2010, SHETTY2014}.  Simple arguments for the KS relation have been presented by several authors \citep{ELMEGREEN1994, WONG2002, KRUMHOLZ2007}, revolving around the idea that a roughly constant fraction of the gas present in molecular clouds will be converted into stars each free-fall time. With $\mathrm{t_{ff}} \propto \rho^{-0.5}$, we would expect $\dot{\rho} \propto \rho^{1.5}$ and with an assumption that scale heights of galaxies do not vary significantly we can come to the conclusion that $\mathrm{\Sigma_{SFR} \propto \Sigma_{gas}^{1.5}}$. 

However the interpretation of the K-S relation is more complicated when we start to consider what ``gas available for star formation'' actually means. The early study by \citet{KENNICUTT1989} used CO emission to trace the gas surface density. Later studies \citep{BIGIEL2010, SHETTY2014} that also focus on CO emission suggested lower values of the K-S index, towards $N \approx  1$ \citep[although this too is under debate:][]{KENNICUTT2012}. 

As one moves to progressively higher density gas tracers, one would expect the correlation between the mass of the gas present and star formation rate to become tighter, provided that the star formation is being measured on time-scales similar to the free-fall time of the gas tracer (see the work by \cite{KRUIJSSEN2014} for a discussion of the time-scales). For example,  \cite{GAO2004a,GAO2004b} conducted a  K-S study using HCN J$\, =\,$1$\, -\,$0\footnote{Note that we will refer to `HCN J$\, =\,$1$\, -\,$0' simply as `HCN' for the rest  of  the  paper} line emission, which is assumed to trace higher density gas than CO, and found that $\mathrm{L_{FIR} \propto L_{HCN}}$. This implies that $\mathrm{\Sigma_{SFR} \propto \Sigma_{gas}}$, and so for HCN emission the K-S index is around 1. \cite{WU2005} suggested that the HCN emission from a galaxy simply counts the number of star-forming clumps present in these galaxies, and so HCN is primarily tracing the densities at which star formation ``sets in''.  This has spawned significant interest in HCN as a tracer of ``dense gas'' in the ISM, and as a tool for studying the star formation relations in more detail. 

However, exactly what density HCN traces is still very much unclear. Although the critical density is quite high --  $n_{\rm crit} = 4.7 \times 10^5 \, \rm cm^{-3}$ for the multi-level definition of the critical density at 10K --  the line is typically optically thick, which can lower the effective critical density, as discussed in detail by \citet{SHIRLEY2015}. Indeed, \citet{SHIRLEY2015}, using simple one-dimensional radiative transfer, demonstrates that a $\mathrm{1\ K\, km\, s^{-1}}$ line can be produced by densities as low as $\mathrm{8.4 \times 10^3\ cm^{-3}}$, due to radiative trapping. This is below the density that HCN was assumed to trace in the studies of  \cite{GAO2004a,GAO2004b}, where they calculated that HCN emission was probing a characteristic density of $\mathrm{3 \times 10^4\ cm^{-3}}$.

\cite{KRUMHOLZ2007a} investigated how the KS law changes with differing molecular gas tracers, including HCN \citep{NGUYEN1992, GAO2004a, GAO2004b,  RIECHERS2006, GAO2007}. Similar to  \cite{GAO2004b}, their model uses an LVG calculation but with the inclusion of a lognormal PDF for the density in the molecular gas in their model.  With this approach they found that HCN emission generally traces dense gas, $\mathrm{n_{dense} \sim 10^5\ cm^{-3}}$.  Their model showed a strong correlation with the observed data with a direct proportionality between far infrared luminosity and HCN luminosity. 

More recently, \citealt{LEROY2017} have used LVG calculations to explore the influence of the density PDF on the characteristic density traced by HCN emission. They found that the characteristic density is highly sensitive to what one assumes regarding the cloud density PDF, with values from their models ranging from $\mathrm{\sim 10^{3} \: cm^{-3}}$ to more than $\mathrm{10^{5} \: cm^{-3}}$. However, one weakness of this and the other simple models described above is that they incorporate little or no information on the spatial distribution of the dense gas, which potentially has a large impact on the relation between HCN optical depth and gas density. A first attempt to properly account for the spatial structure of the dense gas was made by \citealt{ONUS2018}, who post-processed a high resolution simulation of a small portion of a molecular cloud by \citealt{FEDERRATH2015}. Based on this calculation, they predicted that HCN emission traces gas with a luminosity-weighted mean density of $\mathrm{0.8 - 1.7 \times 10^4\, cm^{-3}}$.  

There is also increasing evidence from observational studies of giant molecular clouds (GMCs) in the Milky Way that HCN is probing lower densities than previous assumed.  \cite{PETY2017}, \cite{KAUFFMANN2017} and \cite{BARNES2020} have shown that HCN also traces diffuse regions of molecular clouds at a density of $\sim 500\, \mathrm{cm^{-3}}$, $\sim 10^3\, \mathrm{cm^{-3}}$ and $\mathrm{\sim 10^3\, cm^{-3}}$, respectively. \citet{TAFALLA2021} also show that HCN emission can be detected at visual extinctions ($\mathrm{A_V}$) as low as $\mathrm{\sim 1\, mag}$.

In this paper, we will expand upon the work of \cite{ONUS2018}. First, rather than simulating a small sub-region within a cloud, we will use simulations of low-density 'cloud-cloud collisions', to create dense molecular regions with self-consistent density and velocity fields. Second we will use a detailed model of the heating and cooling processes that is coupled to a time-dependent chemical network that follows H$_2$ and CO formation and destruction.  Although we do not follow the HCN chemistry self-consistently in our study, we will use the results from \citet{FUENTE2019} to relate the HCN abundance at each point in the simulation volume to the CO abundance and the local visual extinction -- two properties that are followed self-consistently in our simulations. We perform radiative transfer (RT) post-processing on the simulations with the publicly available code RADMC-3D \citep{DULLEMOND2012} to make synthetic observations of the HCN (1-0) line, and we use these to explore the density regime traced by HCN emission. Where possible, we have compared to the recent observational studies. 

In Section \ref{sec:numerics}, we describe in detail the magnetohydrodynamical simulations that form the basis of this study, and the radiative transfer post-processing that we perform to get the synthetic HCN emission cubes.  We present the basic evolution of the magnetohydrodynamical simulations in Section \ref{sec:overview}, and discuss how we decide when in the cloud evolution we perform the RT.  The HCN emission is presented in Section \ref{subsec:synthetic}, including the discussion of the density regime that it probes in our simulations.  We relate HCN (1-0) / CO (1-0)  to dense gas in section \ref{subsubsec:relateHCNgas}.
We discuss some possible caveats in our study in Section \ref{sec:disc} and present our conclusions in Section \ref{sec:conc}.  

\section{Numerical Approach}
\label{sec:numerics}
We investigate two spherical clouds that collide head-on at four different velocities using a magnetohydrodynamical (MHD) code that includes a time-dependent chemical network for $\mathrm{H_2}$ and CO formation, which runs alongside a detailed treatment of the heating and cooling in the ISM.   We then post-process our simulations using a synthesised HCN abundance, which is related to the CO abundance in our models (see section \ref{subsec:radmc} for further details), and a radiative transfer code to create HCN emission position-position-velocity (PPV) cubes. These are then analysed to determine the density regime traced by HCN emission. 
\subsection{The Numerical Model}
\label{subsec:model}

We use a modified version of the publicly available moving-mesh code, {\sc {\sc Arepo}} \citep{SPRINGEL2010, WEINBERGER2020}. The adpative, moving mesh in {\sc Arepo} allows us complete control over the resolution in our simulations, while at the same time minimising advection errors. It is thus ideally suited to this type of ISM problem. Our modifications to {\sc Arepo} include: the use of the radiative heating and cooling and cosmic ray heating treatments described by \cite{GLOVER2007, GLOVER2012}; the TREECOL algorithm developed by \cite{CLARK2012TREECOL} to calculate the attenuation of the interstellar radiation field (ISRF); time-dependent chemistry that follows H$_2$ and CO formation (see \citealt{HUNTER2021}); a sink particle algorithm  \citep{BATE1995, FEDERRATH2010} to treat small, gravitationally-collapsing regions associated with star formation \citep{TRESS2020, WOLLENBERG2020}.

\subsection{Initial Conditions}
\label{subsec:Initcond}

Our simulations start with two spherical clouds, each with a radius of $ \mathrm{19.04\ pc}$, a number density of $\mathrm{10\ cm^{-3}}$ (note that we will refer to `number density' simply as `density' for the rest of the paper) and a mass of $\mathrm{1 \times 10^4\ M_{\odot}}$. Both clouds have an initial temperature of $\mathrm{300\ K}$, consistent with the balance between fine structure cooling and photoelectric heating at a number density of $\mathrm{10\ cm^{-3}}$.  The geometry of the simulation is such that the cloud centres are placed at a distance of $\mathrm{57.11\ pc}$ and $\mathrm{114.23\ pc}$ respectively in x, while both centres are placed at a distance of $\mathrm{85.67\ pc}$ in both y and z in a cuboid of size $\mathrm{171.34\ pc}$. The velocity of each cloud is mirrored along x such that they are sent on a collision course with one another. Four different velocities are chosen to cover the typical and extremes of the velocity distribution of the gas flow in spiral arms \citep{DOBBS2008}; $\mathrm{ 1.875\ km\,s^{-1}}$, $\mathrm{3.75\ km\,s^{-1}}$, $\mathrm{7.5\ km\,s^{-1}}$ and $\mathrm{15\ km\,s^{-1}}$ (note that these quoted velocities are the velocities of the individual clouds, i.e. the relative velocity is twice these values). An initial turbulent velocity field is applied to the clouds, which follows a $\mathrm{P(k) \propto k^{-4}}$ scaling law with a natural mix of solenoidal to compressive modes. The velocity dispersion of the turbulence is set to $\mathrm{1.16\ km \,s^{-1}}$, which provides virial balance between the (bulk) kinetic and gravitational energies. By allowing a period of time between the initial set-up and the cloud collision, the supersonic turbulence (Mach number of $\sim 2$) has the chance to create structure in the clouds before they encounter the main collisional shock. 

Each cloud is initially modelled with 2,000,000 cells, randomly generated in a sphere, such that the initial cell mass is $\mathrm{0.005\ M_{\odot}}$. A further 262,144 cells, with mass $\mathrm{0.066\ M_{\odot}}$ are randomly injected into the rest of the computational domain to model the background gas, which is taken to have a density of $\mathrm{0.063\ cm^{-3}}$. As the simulation progresses, the mesh is constantly monitored to maintain a cell mass of roughly $\mathrm{0.005\ M_{\odot}}$. On top of this, we impose three further resolution criteria. The first criterion is that the Jeans length is resolved by at least 16 cells, to make sure we correctly capture the fragmentation in the gas. The second criterion is that the volume of neighbouring cells differs by no more than a factor of 8. Finally, we set a minimum and maximum cell size of $\mathrm{100\ AU}$, and $\mathrm{12\ pc}$, respectively. 

Due to self-gravity in {\sc Arepo}, the gas in our simulations has the ability to form regions of high density that can undergo runaway gravitational collapse. The final outcome of such a process would be the formation of a star or small stellar system.  We employ sink particles \citep{BATE1995} to model these objects, and to follow both their dynamics and further accretion.  In this study, several conditions must be met for a gas cell to be turned into a sink particle, which follows the criteria laid out in   \citet{FEDERRATH2010}. First, the candidate cell must be above our sink creation density $\mathrm{n_{sink} = 10^8\ cm^{-3}}$, and be a local minimum in the gravitational potential.   Then we require that the gas within the sink accretion radius, $\mathrm{r_{sink}}$ -- here taken to be $\mathrm{185\ AU}$ -- must be gravitationally bound, and both moving towards {\em and} accelerating towards the candidate's location -- that is, the mass-weighted $\nabla \cdot {\bf v}$ and $\nabla \cdot {\bf a}$ within the sink creation radius must be negative).  

We make use of the magnetohydrodynamical module in {\sc Arepo}, as described in \cite{PAKMOR2011}. This includes hyperbolic divergence cleaning \citep{DEDNER2002} and the divergence advection terms introduced by \cite{POWELL1999}. In our initial setup, we include a uniform magnetic field of strength $\mathrm{3\ \mu G}$, directed along x such that the collision is occurring along the magnetic field lines. This value of  $\mathrm{3\ \mu G}$ is consistent with that found from observations of clouds with number densities similar to those we study here \citep{CRUTCHER2010, CRUTCHER2012}. 

We adopt a composition characteristic of the local ISM for the metals and dust included in our ISM model (i.e. the heating and cooling and time-dependent chemistry).  The initial abundances of carbon and oxygen are set to $x_\mathrm{C} = 1.4 \times 10^{-4}$ and $x_\mathrm{O} = 3.2 \times 10^{-4}$, respectively, as given by \cite{SEMBACH2000}, where $x_i$ is the fractional abundance of the element relative to hydrogen nuclei. We assume that the carbon starts in the form of C$^+$ due to the photoionization by the ISRF, and that oxygen starts in the form of neutral oxygen. We also assume that most of the hydrogen starts in atomic form, and that a small amount of hydrogen is in H$^+$ form ($x_{\mathrm{H^+}} \sim 0.01$) due to the balance of cosmic ray ionization and recombination.

For our model of the interstellar radiation field (ISRF), we adopt the spectral shape described in \cite{MATHIS1983} at longer wavelengths and \cite{DRAINE1978} at UV wavelengths. The strength of the ISRF is $\mathrm{G_0\ =\ 1.7}$ in \cite{HABING1968} units (see \citealt{DRAINE2011}) and the cosmic ray ionization rate of atomic hydrogen is set to $\mathrm{\zeta_H\ =\ 3\ \times 10^{-17}\ s^{-1}}$.

\subsection{Radiative transfer post-processing}
\label{subsec:radmc}

\begin{figure}
    \includegraphics[width=80mm]{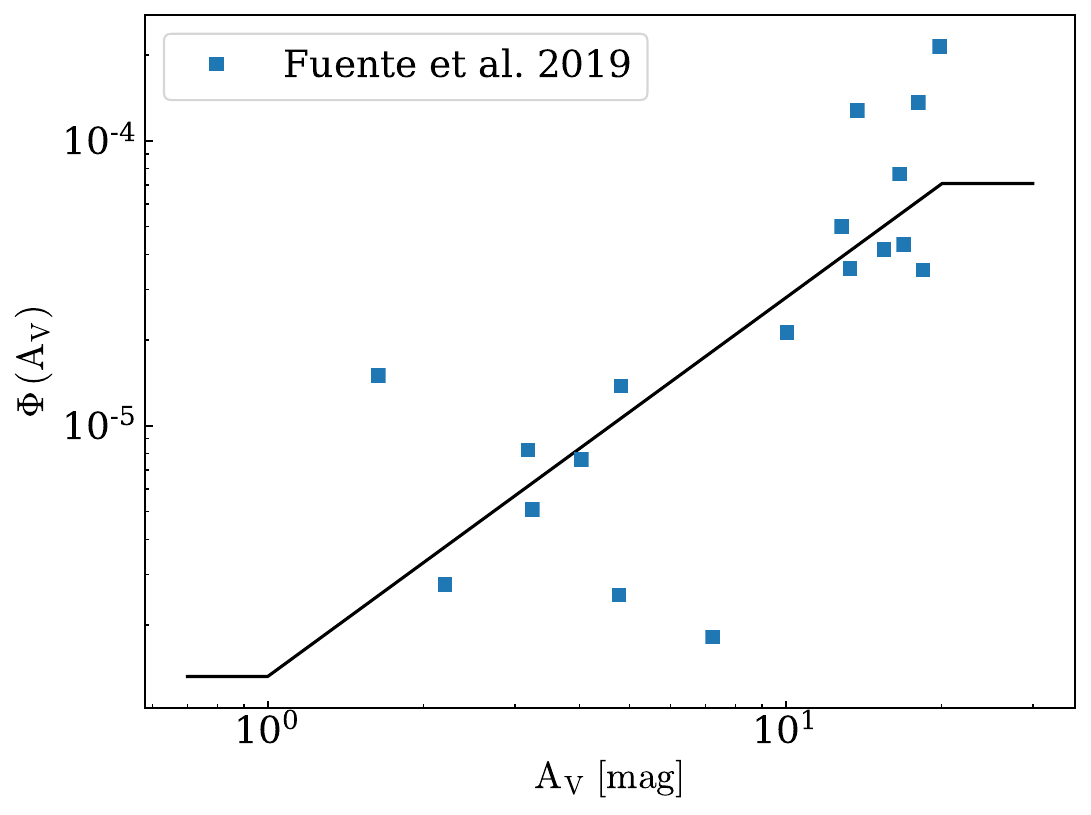}
    \caption{Our conversion factor from CO abundance to HCN abundance, $\Phi\, \mathrm{(A_V)}$ (black line), together with the observational measurements of the HCN to CO ratio presented in \citealt{FUENTE2019} (blue squares)}   
    \label{fig:xhcnav}
\end{figure}

We use the RADMC-3D radiative transfer (RT) code \citep{DULLEMOND2012} to create post-processed position-position-velocity (PPV) cubes of HCN emission from our {\sc Arepo} simulations.  We make use of internal functions in {\sc Arepo} to create a regular cartesian grid of fluid properties that can be converted to a form that is compatible with the fixed cartesian grid used by RADMC-3D. This means however, that we cannot perform the RT post-processing on the entire computational domain that is evolved in {\sc Arepo}.  We therefore limit our RT analysis to a $\mathrm{10\ pc}$ cubic region that envelopes the highest density region in the cloud-cloud collision; {\sc Arepo}'s voronoi mesh is interpolated on to a $450^3$ grid such that we have a spatial resolution of $\mathrm{0.022\ pc}$ in the RT. This is sufficient to capture both the scales of the molecular cloud and the cores that form within. 

\begin{figure*}
\centerline{
    \includegraphics[width=200mm]{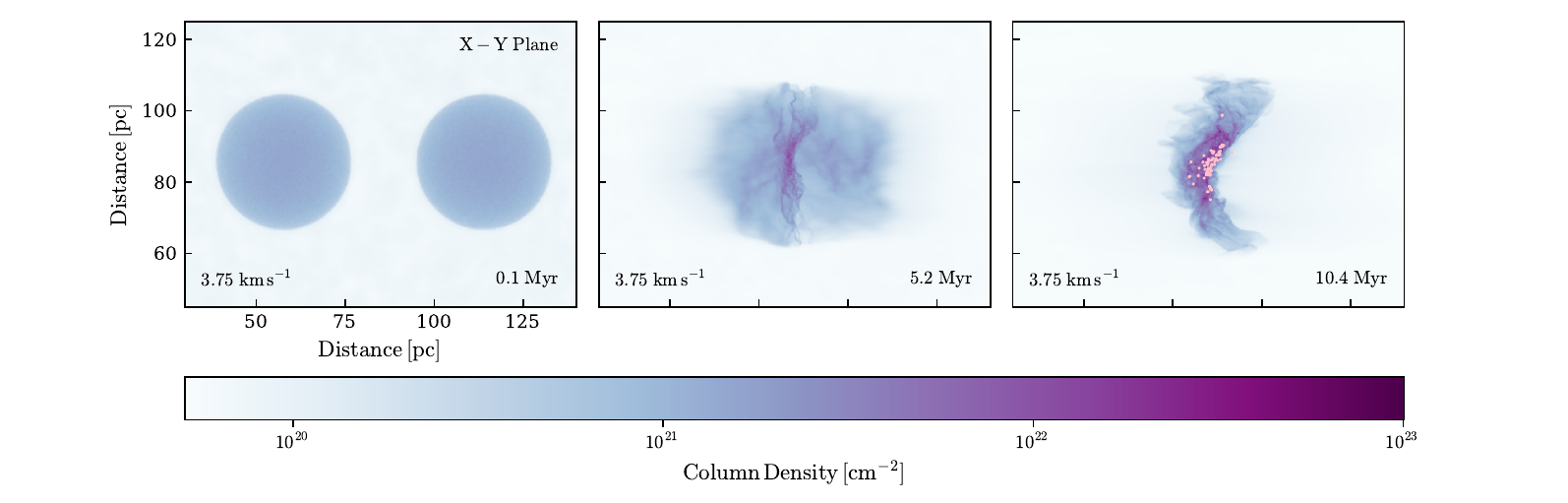}
    }
    \caption{Column density of the 3.75 $\mathrm{km\, s^{-1}}$ simulation at three different times, one at the start of our simulation, and the other two at 5.2 Myr intervals. The sink particle locations are also included in any image that possess sink particles which are represented as pink points.}
    \label{fig:initfig} 
\end{figure*}

Due to the complex nature of nitrogen chemistry, it is currently computationally intractable to self-consistently compute the time-dependent abundance of HCN in our {\sc Arepo} chemical network.  We therefore make use of the observationally-derived $x_{\mathrm{HCN}}/x_{\mathrm{CO}}$ v $\mathrm{A_V}$ relation shown in Figure 8 of \cite{FUENTE2019}. Note that other interpretations of the data in Fuente et al. (2019) are possible. For example from their Figure 6, one could infer that $x_{\rm HCN}$ is proportional to $\mathrm{A_V}$ above $\mathrm{A_V} \approx 10$, and flat below this. However by tying the HCN abundance to the CO abundance, we can capture (phenomenologically) the effects of the photo-destruction of the HCN by the ISRF, and thus avoid spuriously large HCN abundances at low densities and A$_{\rm V}$.

When using the $x_{\mathrm{HCN}}/x_{\mathrm{CO}}$ v $\mathrm{A_V}$ relation in Fuente et al (2019), our visual extinction is calculated along the same line of sight as the rays used to solve the RT problem in RADMC-3D, simply by first getting the column density at each point via,
\begin{equation}
N_{\rm H} = \sum_{i = 0}^{N_{\rm LoS}} \frac{\rho_i}{1.4 \, \rm m_{p}} \, \Delta L    
\end{equation}
where $i$ denotes each of the $N_{\rm LoS}$ cells along the line-of-sight, with densities $\rho_i$ and length $\Delta L$; the term $1.4 \rm m_{p}$ converts to the number density of hydrogen nuclei. We then convert this to a visual extinction via,
\begin{equation}
\label{eq:AV} 
\mathrm {A_V  = \frac{N_H} {1.87 \times 10^{21} \rm cm^{-2}}}
\end{equation}
where here ${\rm N_H}$ is the column number density of hydrogen {\em nuclei} \citep{BOHLIN1978, DRAINE1996}. The column density thus derived is designed to mimic the observed column density used in Figure 8 of \cite{FUENTE2019}. Note that this differs from the column density as seen by each cell in our simulation, which is derived via our {\sc TREECOL} algorithm.

Using this relationship, our computed CO abundances derived using {\sc Arepo} can then be used to calculate our HCN abundance with respect to hydrogen nuclei. We then compute the HCN abundance in each RADMC-3D grid cell via,
\begin{equation}
\label{eq:xHCN}
x_{\mathrm{HCN}} = \Phi\, \mathrm{(A_V)} \cdot x_{\mathrm{CO}}
\end{equation}
where $\Phi\, \mathrm{(A_V)}$ is our conversion factor from CO abundance to HCN abundance, $x_{\mathrm{HCN}}$ is the abundance of HCN relative to hydrogen nuclei and $x_{\mathrm{CO}}$ is the abundance of CO also relative to hydrogen nuclei. The value of $\Phi\, \mathrm{(A_V)}$ is obtained from the \citet{FUENTE2019} results, and we present the data used in Figure \ref{fig:xhcnav}. Note that the data in \citet{FUENTE2019} covers a limited range in $\mathrm{A_V}$. Rather than make up a relation outside these limits, we simply hold the conversion factor constant with increasing/decreasing $\mathrm{A_V}$. While one might expect this to cause problems at low $\mathrm{A_V}$ -- potentially boosting the HCN abundance -- in practise this does not happen, as at low $\mathrm{A_V}$ the CO abundance in any case self-consistently falls to zero due to our treatment of the photodissociation. Note that another caveat in our model is that the HCN formation timescale is assumed to be exactly equivalent to the CO formation timescale. Although this is unlikely to be exactly the case, the recent results from \cite{PRIESTLEY2021} -- which captured the non-equilibrium chemistry of a dynamically evolving cloud --  demonstrate that both the CO and HCN formation timescales are shorter than the dynamical timescale in scenarios similar to those we study here. Thus our coupling of the HCN abundance to the CO abundance is unlikely to affect the results. \cite{PRIESTLEY2020} found that a large variation in the distribution of the HCN abundance in the density space leads to very little variation in the intensity of HCN.

The level populations of HCN were calculated in RADMC-3D using the large velocity gradient (LVG) approximation \citep{SOBOLEV1957} as implemented by \cite{SHETTY2011}. We use the collisional rate data for HCN provided by Leiden Atomic and Molecular Database \citep{SCHOIER2005, FAURE2007, DUMOUCHEL2010}. In this study, we use the version of the HCN line data without hyperfine structure, and we include excitation from two collisional partners, $\mathrm{H_2}$ and electrons. The radiative transfer is performed along the $z$-axis of the grid (perpendicular to the axis of the cloud-cloud collision), such that the rays are directed from negative to positive $z$.



%
%
%
\section{Overview of the cloud-cloud collision simulations}
\label{sec:overview}
We present four cloud-cloud collision simulations, with each investigating a different collision velocity, as outlined in Section \ref{subsec:Initcond}.  In Figure \ref{fig:initfig}, we show the evolution of the 3.75 $\mathrm{km\, s^{-1}}$ simulation from its initial conditions, to give the reader a better understanding of how our clouds evolve.  Figure \ref{fig:initfig} shows that even though the simulations begin from unrealistic spheres, the simulations evolve over time to form dense filamentary structures, consistent with the chaotic environment familiar from both previous colliding flow models and from observational studies of molecular clouds. As the clouds meet, the two supersonic colliding flows cause a shocked layer at the point of impact creating a layer of dense gas. This process repeats as more of the inflowing gas from the opposing edges of the clouds fall into the shocked dense region.

All simulations are evolved to a point $\mathrm{\sim 1 - 3\ Myr}$ after the formation of the first sink particle, which we will denote as $\mathrm{t_ {SF}}$ -- the time of ``star formation''.  In each of simulations, $\mathrm{t_{SF}}$ occurs at roughly 15 Myr, 11 Myr, 9 Myr and 12 Myr for the 1.875 $\mathrm{km\, s^{-1}}$, 3.75 $\mathrm{km\, s^{-1}}$, 7.5 $\mathrm{km\, s^{-1}}$ and 15 $\mathrm{km\, s^{-1}}$ initial bulk velocities respectively. 

%
%
\begin{figure*}
    \centerline{
        \includegraphics[width=85mm]{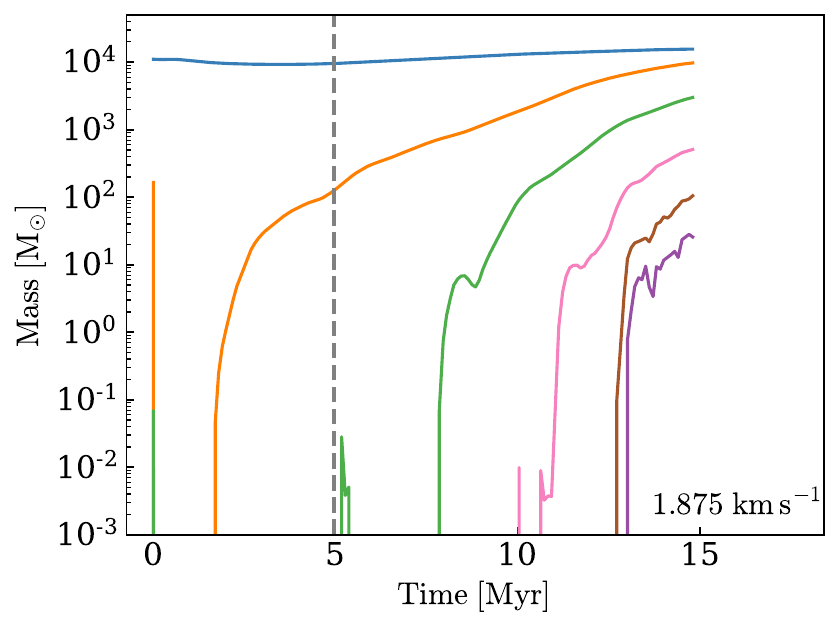}   \includegraphics[width=85mm]{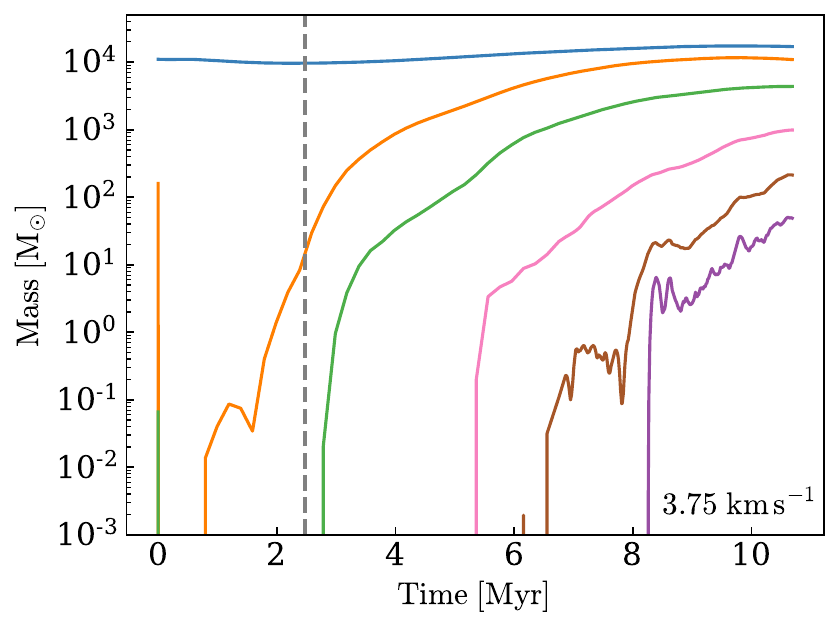}
    }
    \centerline{    
        \includegraphics[width=85mm]{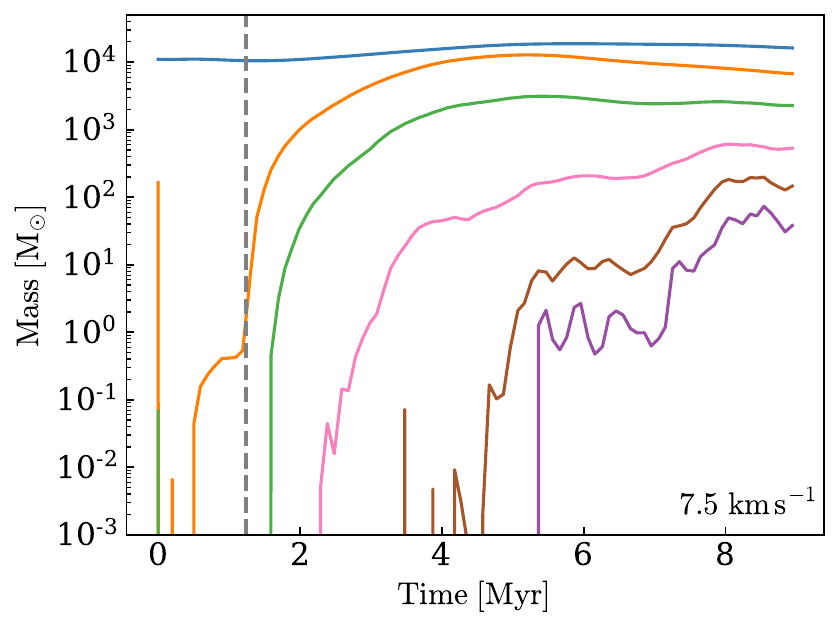}         \includegraphics[width=85mm]{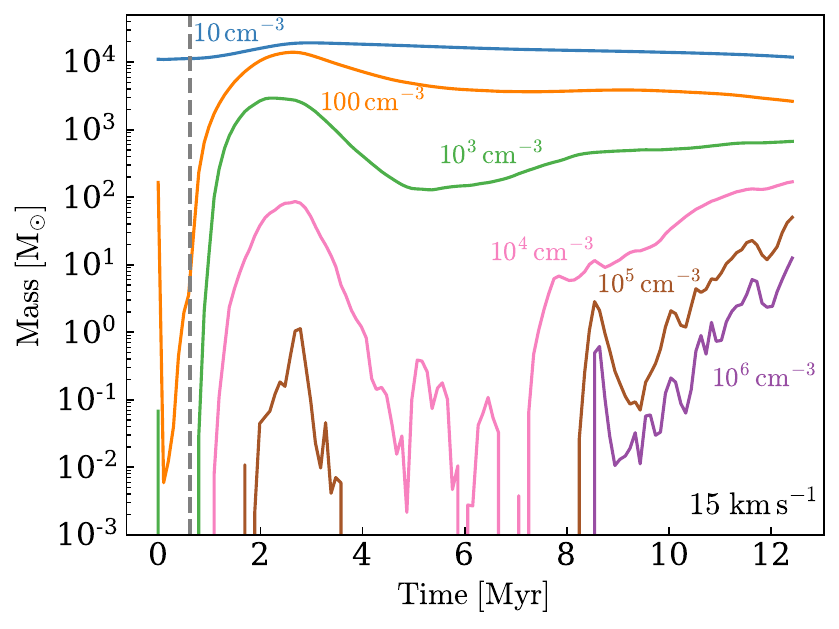}
    }
    \caption{Evolution of mass of the gas within the simulations that lies above a given density threshold, which is labelled on the final plot. Each plot shows the evolution of mass with different initial velocities. From left to right, we have initial velocities of 1.875 $\mathrm{km\, s^{-1}}$, 3.75 $\mathrm{km\, s^{-1}}$, 7.5 $\mathrm{km\, s^{-1}}$, 15 $\mathrm{km\, s^{-1}}$. Note that the gray, vertical line denotes the approximate time at which the edge of both clouds come into contact.}
    \label{fig:evolmass}
\end{figure*}

In Figure \ref{fig:evolmass}, we plot the evolution of the mass above different density thresholds ($\mathrm{n_{thr}}$) as a function of time. We see that a higher collision velocity between the two clouds decreases the time taken to form gas above $\mathrm{n_{thr}} \sim$ $10^4\ \mathrm{cm^{-3}}$. However this trend does not continue as we move to higher $\mathrm{n_{thr}}$: the faster flow of 15 $\mathrm{km\, s^{-1}}$ has a clear difficulty in forming gas with density above $10^6\ \mathrm{cm^{-3}}$, and indeed actively seems to lose gas above $\mathrm{n_{thr}} \sim 10^4 \rm \, cm^{-3}$ at times between $\mathrm{\sim 4 - 8\ Myr}$. This implies that much of the dense gas that is initially created in the 15 km s$^{-1}$ simulation is not self-gravitating, and either re-expands once the confining flow has finished, or is shredded by further interactions with surrounding flows.  Only at  late times (beyond around 8 Myr), once some of the collisional kinetic energy has been dissipated, are gravitationally bound regions able to form, providing an increase in the dense gas fractions. The dip and rise in the $\mathrm{n_{thr}} = 10^3 \, \rm cm^{-3}$ line indicates that this occurs at initially quite low densities (and thus large scales). 

Using Figure \ref{fig:evolmass}, we can see that by the point at which we terminate the simulation (as presented in the graph's timeline), somewhere between 0.5-3.6\% of the total cloud mass sits above a density of $10^4\ \mathrm{cm^{-3}}$ and 0.04-0.2\% of the total cloud mass above a density of $10^6\ \mathrm{cm^{-3}}$.  The simulations have therefore evolved far enough for us to proceed with the analysis of the HCN emission with RADMC-3D, as they contain gas at densities commonly associated with prestellar cores, and starting to form sink particles.  

\begin{figure*}
    \centerline{
        \hspace{0.7cm}
	    \includegraphics[width=185mm]{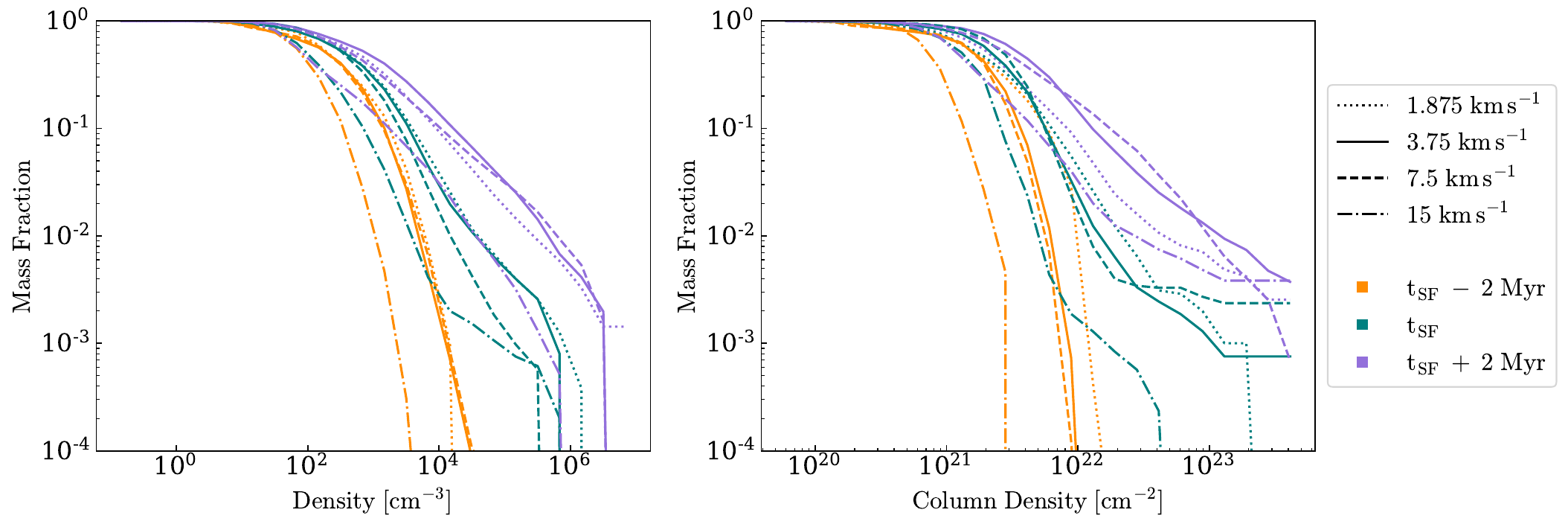}
    }
    \caption{Comparison of the normalized mass-weighted complementary cumulative distribution function (CCDF) for both density and column density for all four simulations at the three different output times. Note that we vary our line-styles based on the initial cloud velocities at the start of our simulations and vary the colour based on the three different output times.}
    \label{fig:cdf}
\end{figure*}

Before looking at the HCN emission, we first examine how much `dense' gas is actually present in the simulations. We show in Figure \ref{fig:cdf} a normalized mass-weighted complementary cumulative distribution function (CCDF) for both density and column density for all twelve simulations. It is clear from Figure \ref{fig:cdf} that the amount of high density gas is more sensitive to when we look at the simulation, than the collision velocity of the flow. Again, we see the same behaviour as in Figure \ref{fig:evolmass}, in that the fastest flow actually has (typically) the least amount of dense gas. With the other flows, the situation is more complicated. It is also clear that as stated previously, there is only a very small percentage of a molecular cloud's mass that goes into these high volumetric densities and column densities.   

\begin{figure*}
    \centerline{
    \includegraphics[width=200mm]{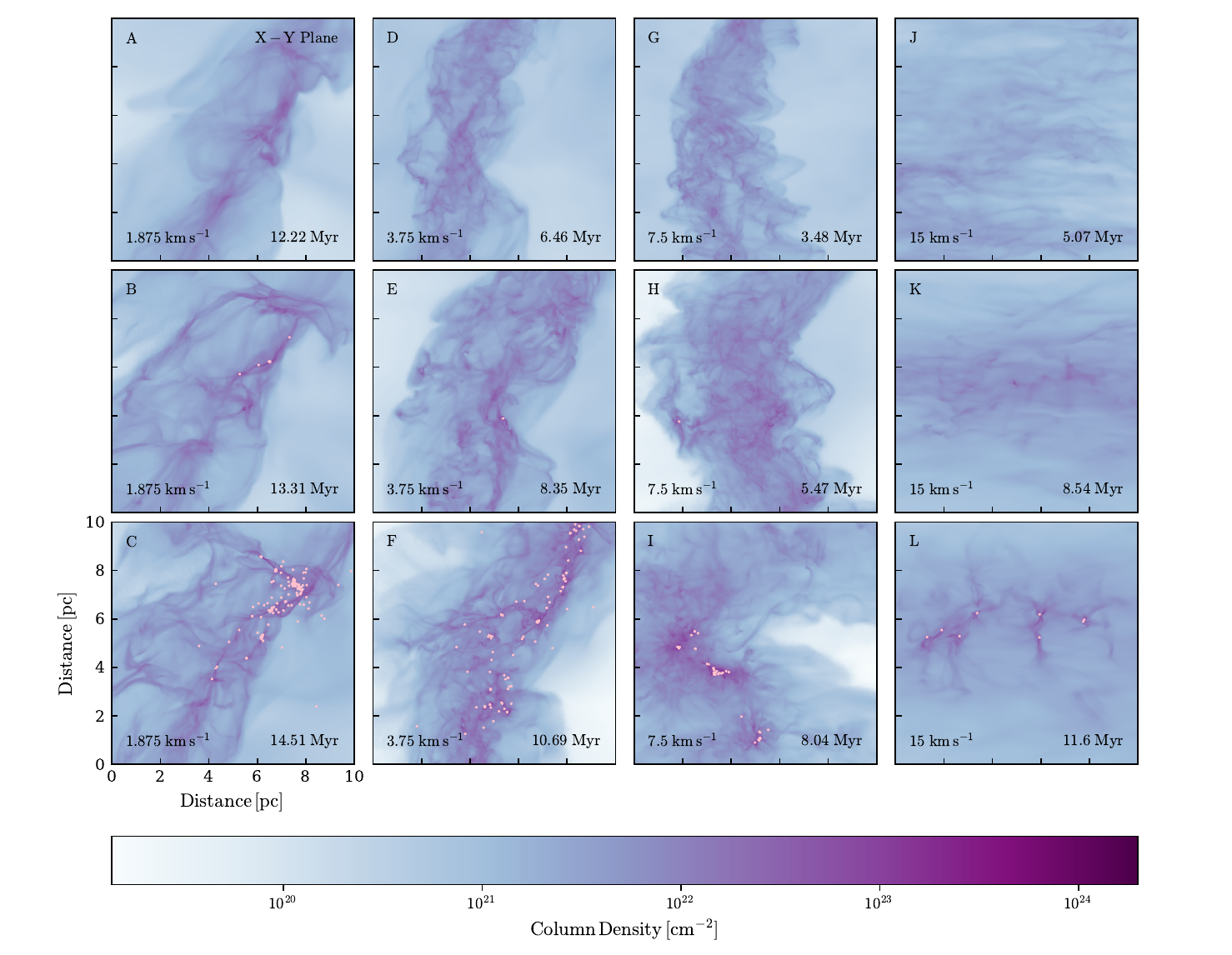}}
    \caption{This grid of column densities shows the state of each simulation at the three points in their evolution where the RT is performed. The sink particle locations are also included in any image that possess sink particles which are represented as pink points. Note that the alphabetical letters on each tile corresponds to the IDs in Table \ref{tab:tableoverview}.}    
    \label{fig:gridcol}
\end{figure*}

To investigate how the HCN emission varies with time, we perform a RT analysis at three different times in each simulation. The first time is taken to be $\mathrm{t_{SF}}$, and then we take two further times, roughly $\mathrm{\sim 2\, Myr}$ before and after $\mathrm{t_{SF}}$. In total this provides twelve different RT simulations of the HCN emission arising from a region of gas $10\, \mathrm{pc}$ on a side. These regions have fully consistent density, velocity and magnetic fields, and have thermodynamics set by our detailed treatment of the heating and cooling in molecular clouds.  The details of these regions are tabulated in Table \ref{tab:tableoverview}, and in Figure \ref{fig:gridcol} we show the column density maps of each of the twelve cases that are used as the initial conditions for our post-processing with RADMC-3D.

%

%
%

%
%
\begin{figure*}
    \centerline{
    \includegraphics[width=200mm]{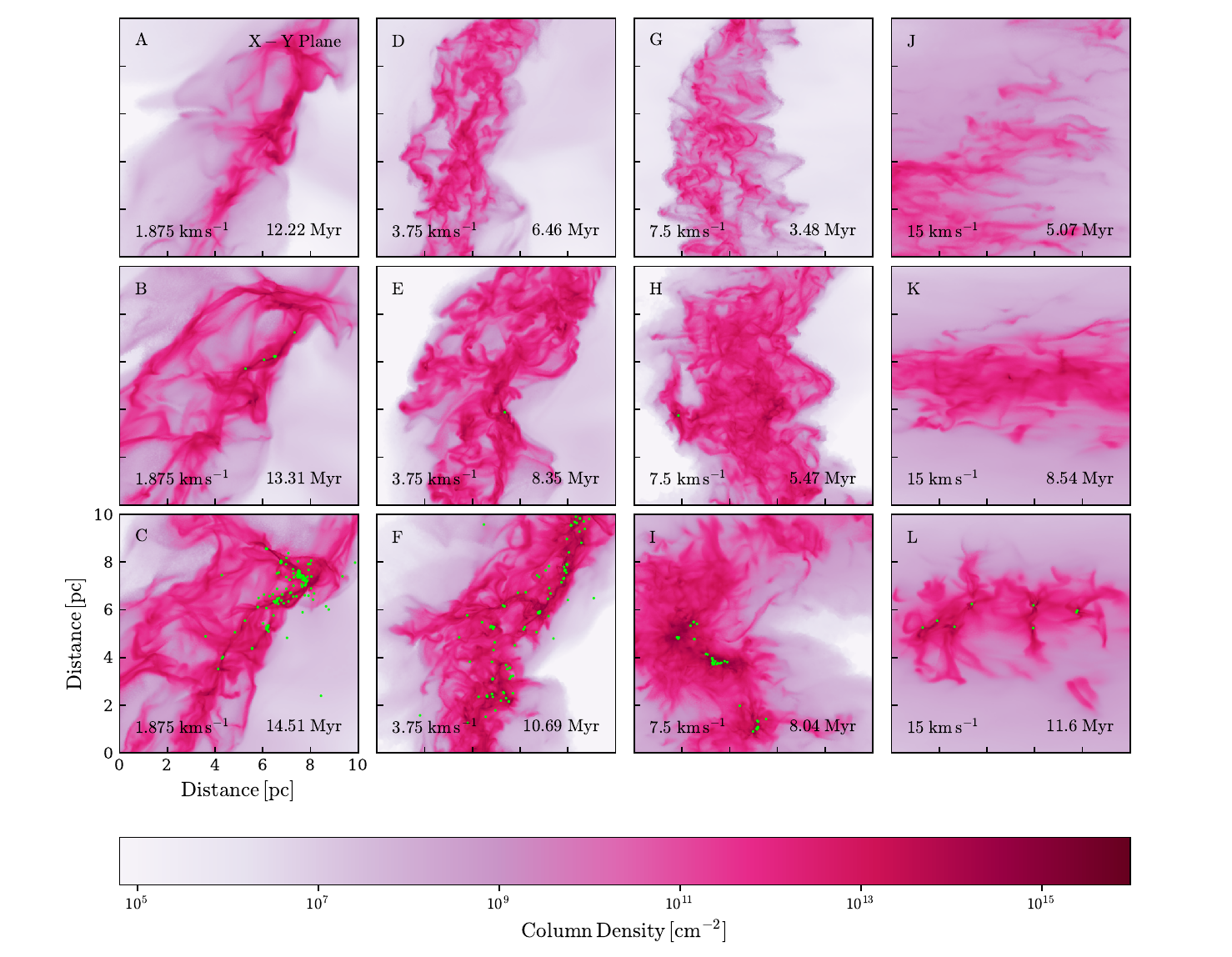}}
    \caption{A grid of Column densities of HCN for all simulations presented in the same format as in Figure \ref{fig:gridcol}.}    
    \label{fig:gridcolHCN}
\end{figure*}

%
%
\begin{figure*}
    \centerline{
    \includegraphics[width=200mm]{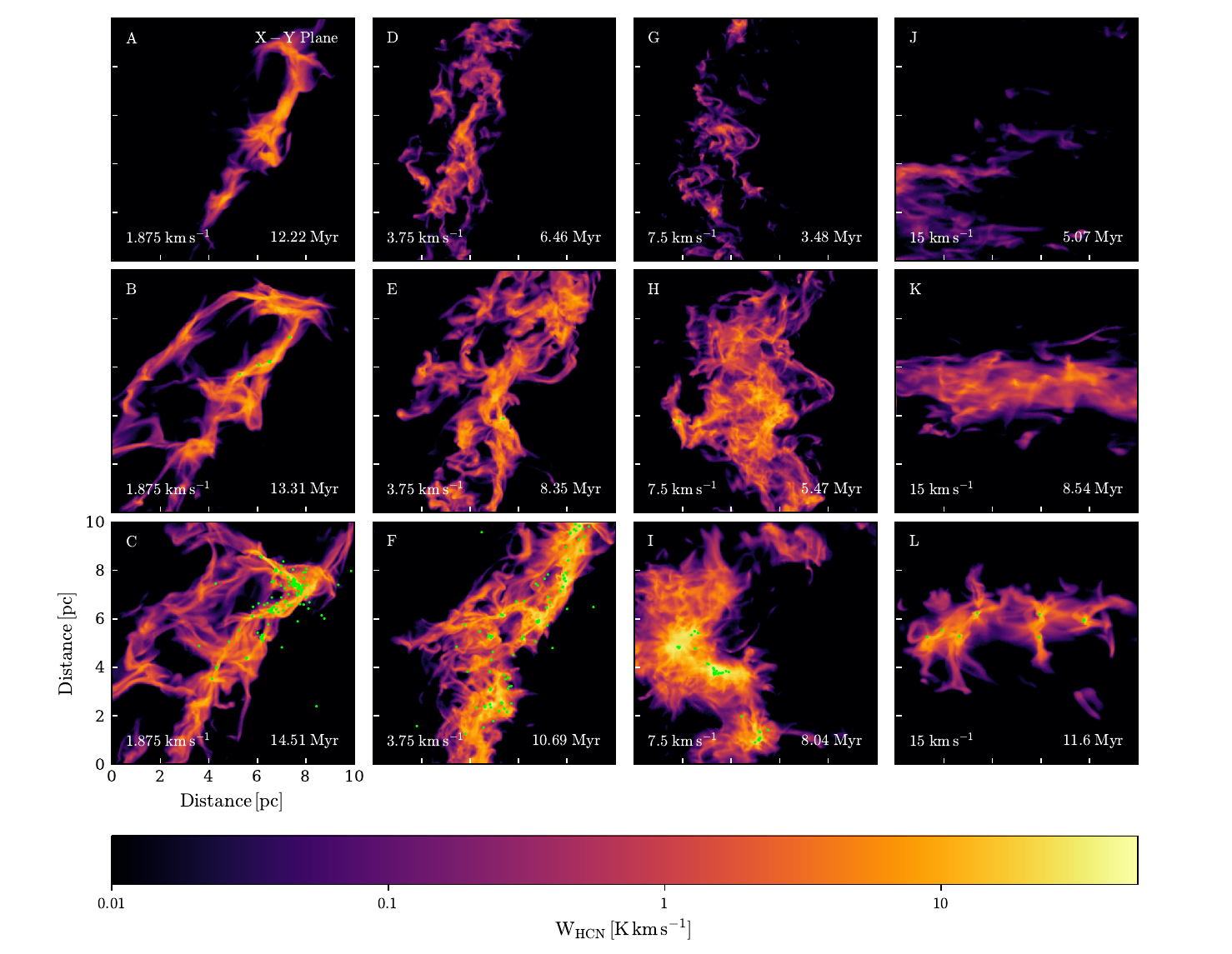}}
    \caption{Maps of the velocity-integrated intensity in the J = 1 $\rightarrow$ 0 line of HCN, $\mathrm{W_{HCN}}$, at the three simulation times and the four initial velocities that were used in Figure \ref{fig:gridcol}. Note that the alphabetical letters on each tile corresponds to the IDs in Table \ref{tab:tableoverview}.}    
    \label{fig:ivmap}
\end{figure*}

%
\begin{figure}
    \includegraphics[width=80mm]{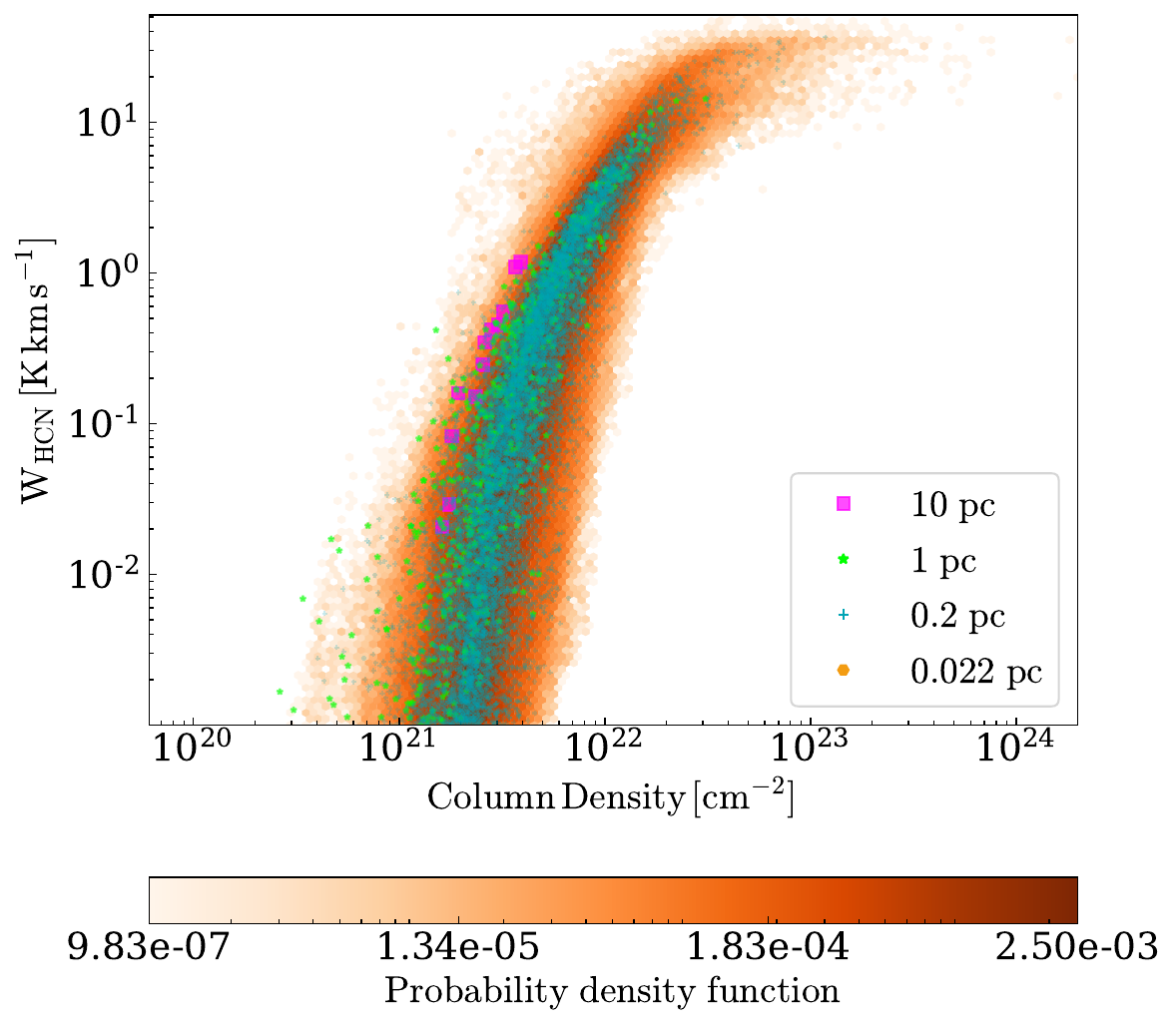}
    \caption{Velocity-integrated intensity in the J = 1 $\rightarrow$ 0 line of HCN, $\mathrm{W_{HCN}}$, plotted against the column density for all simulations collated into one figure. We also demonstrate what happens if we degrade the spatial resolution of the PPV cubes to 0.2, 1 or 10 pc.}
    \label{fig:NH2W}
\end{figure}

%
%
\begin{figure}
	\includegraphics[width=80mm]{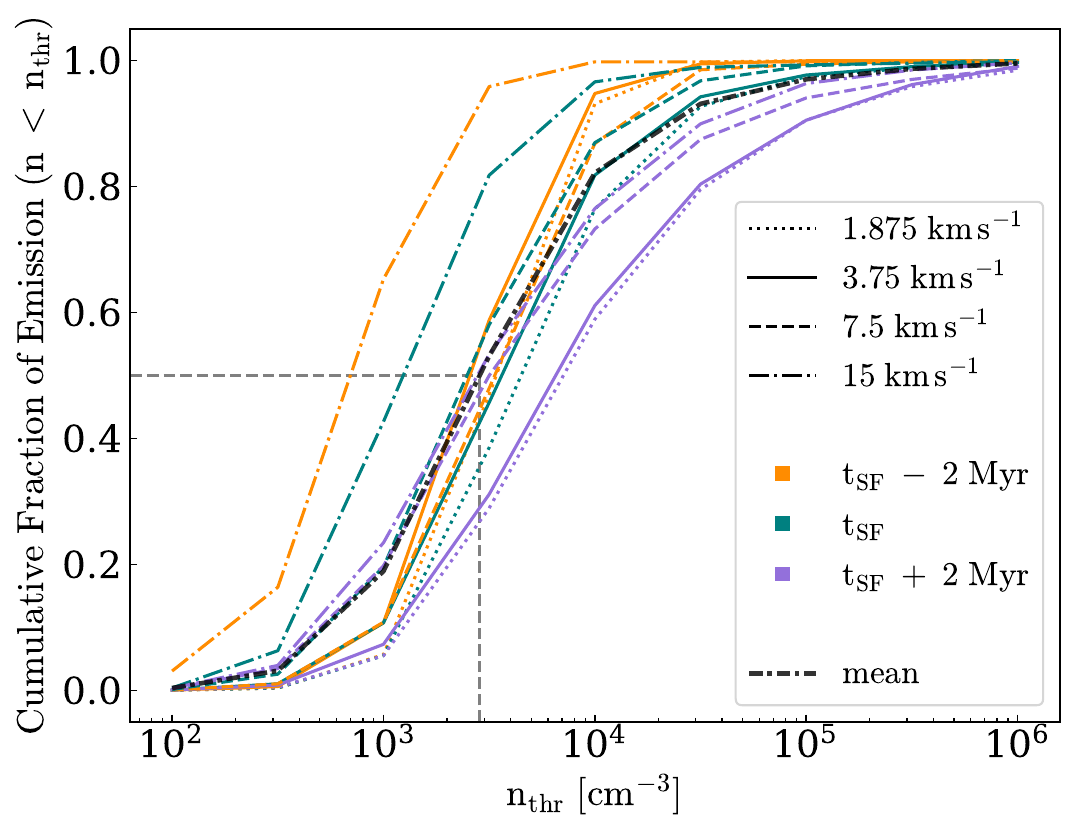}
    \caption{Cumulative fraction of emission of HCN plotted as a function of density for all four simulations at the three different output times. Note that we use the same stylistic format as Figure \ref{fig:cdf}. Also included is the mean cumulative fraction of emission (densely dashdotted black line), plotted as a function of density. The dashed grey line indicates the mean value of the characteristic density, $\mathrm{n_{char} = 2.9 \times  10^{3} \: cm^{-3}}$.}
    \label{fig:fracemall}
\end{figure}

\begin{figure}
	\includegraphics[width=80mm]{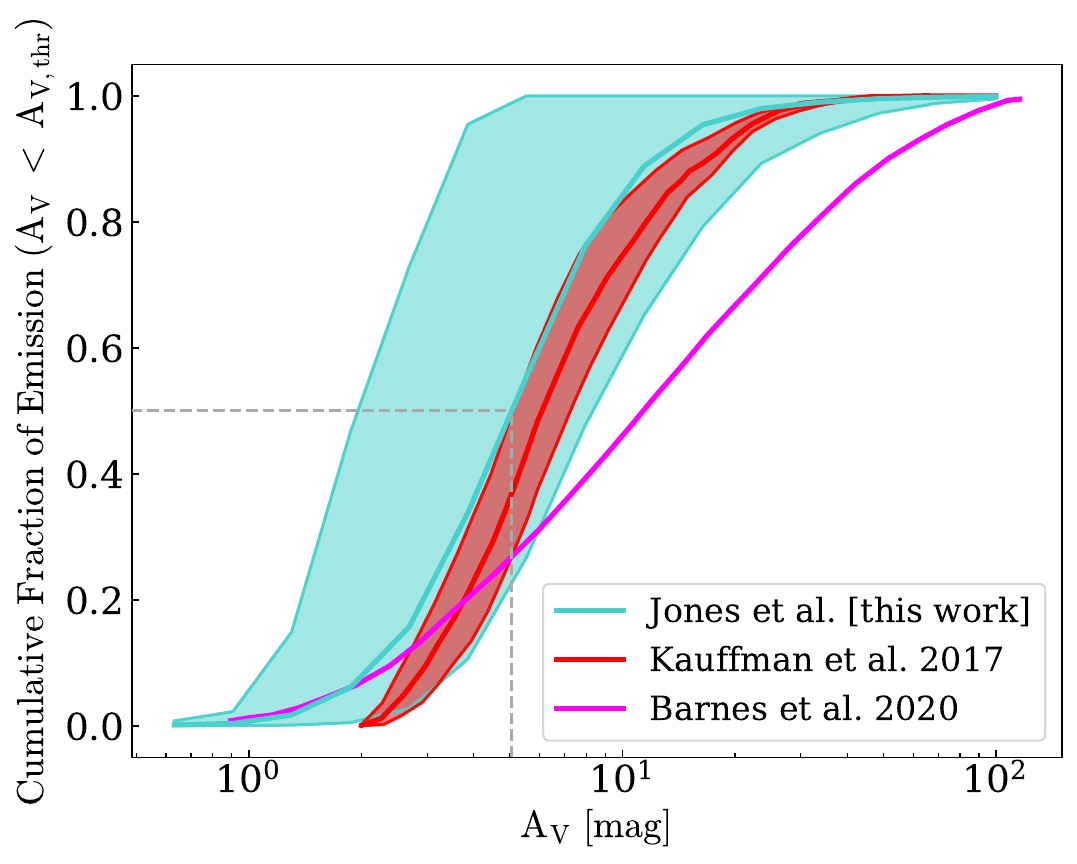}
    \caption{Cumulative emission of HCN plotted as a function of $\mathrm{A_V}$ with the mean, minimum and maximum cumulative emission from all twelve simulations. We include the cumulative emission of HCN from \citet{KAUFFMANN2017} and \citet{BARNES2020} for comparison.}
    \label{fig:fracemav}
\end{figure}

%
%

%
%

%
%
\begin{figure}
    \includegraphics[width=80mm]{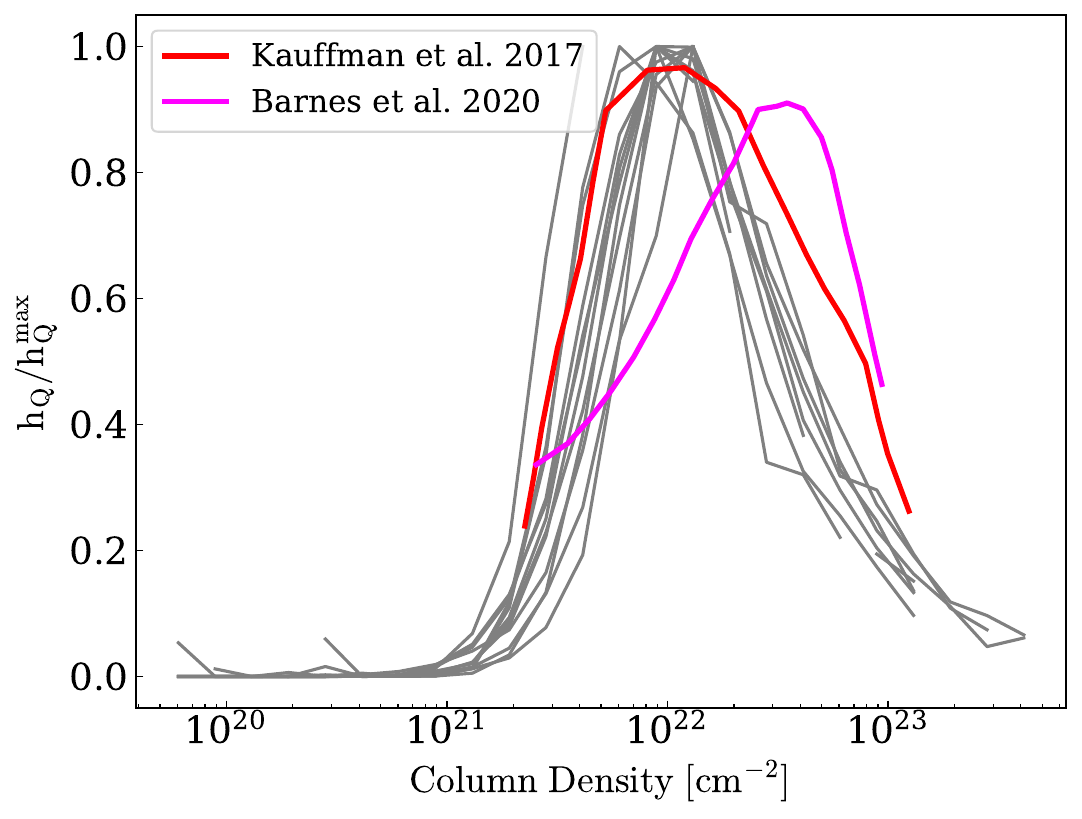}
    \caption{A normalized ratio of the integrated intensity to the column density, $\mathrm{h_{HCN}/h_{HCN}^{max}}$, where $\mathrm{h_{HCN}\ =\ W_{HCN}/N_{H_2}}$, as a function of column density. All of our twelve simulations are grey for clarity. We include \citet{KAUFFMANN2017} and \citet{BARNES2020} for comparison.}    
    \label{fig:h}
\end{figure}

\section{HCN emission from our clouds}
\label{subsec:synthetic}

\subsection{The density regime probed by HCN emission}
\label{subsubsec:HCNdensity}

%
%
\begin{table*}
	\centering
	\caption{A brief overview of all twelve cases that were post-processed through RADMC-3D.}
	\label{tab:tableoverview}
	\begin{tabular}{lcccccccccr} 
		\hline
		& ID & Initial velocity & Time & Mass & HCN luminosity & $\mathrm{W_{HCN}^{mean}}$ & \multicolumn{2}{c}{Fraction of mass above} & \\
		& & & & & & & $\mathrm{2.85 \times 10^{3}\ cm^{-3}}$ & $\mathrm{3 \times 10^{4}\ cm^{-3}}$ & \\
		& & $\mathrm{[km\, s^{-1}]}$ & $\mathrm{[Myr]}$ & $\mathrm{[M_\odot]}$ & $\mathrm{[K\, km\, s^{-1}\, pc^2]}$ & $\mathrm{[K\, km\, s^{-1}]}$ & $\mathrm{[\%]}$ & $\mathrm{[\%]}$ & \\
		\hline
		& A & 1.875 & 12.22 & 2176.49 & 15.92 & 0.16 & 9.37 & 0.02 & \\
		& B & 1.875 & 13.31 & 2929.12 & 34.67 & 0.35 & 19.36 & 1.78 & \\
		& C & 1.875 & 14.51 & 3590.91 & 55.43 & 0.55 & 27.11 & 5.69 & \\
		& D & 3.75 & 6.46 & 2026.82 & 8.28 & 0.08 & 6.65 & 0.03 & \\
		& E & 3.75 & 8.35 & 3153.03 & 42.40 & 0.42 & 18.18 & 1.52 & \\
		& F & 3.75 & 10.69 & 4129.07 & 109.79 & 1.09 & 35.23 & 8.69 & \\
		& G & 7.5 & 3.48 & 1971.21 & 2.93 & 0.03 & 6.61 & 0.03 & \\
		& H & 7.5 & 5.47 & 3414.21 & 45.39 & 0.45 & 13.71 & 0.69 & \\
		& I & 7.5 & 8.04 & 4380.66 & 118.16 & 1.18 & 25.28 & 6.92 & \\
		& J & 15 & 5.07 & 1817.20 & 2.14 & 0.02 & 0.19 & 0.00 & \\
		& K & 15 & 8.54 & 2636.88 & 15.15 & 0.15 & 2.82 & 0.17 & \\
		& L & 15 & 11.60 & 2867.27 & 24.60 & 0.25 & 9.35 & 1.70 & \\
		\hline
	\end{tabular}
\end{table*}

To get a better sense of the formation of the HCN in our simulations, we present a HCN column density in Figure \ref{fig:gridcolHCN} that is from the resulting application of Equation \ref{eq:xHCN} to our CO data in our simulation to generate our HCN abundances combined with Figure \ref{fig:gridcol}.

Using the post-processed HCN abundances generated from {\sc Arepo}, we create PPV cubes of the HCN (J = 1 $\rightarrow$ 0) line emission using RADMC-3D for all twelve regions (three different times for each of our four simulations).  All RADMC-3D simulations track the spectrum between $\mathrm{-3\ to\ 3\ km\, s^{-1}}$ in $\mathrm{0.02\ km\, s^{-1}}$ increments; this is sufficient to completely cover the range of velocities along the $z$ direction in each of our 10pc boxes, while allowing us to model the thermal line-width with around ten points. 

From these PPV cubes, we create velocity-integrated intensity maps of HCN, which can be seen in Figure \ref{fig:ivmap}. The contribution of each column density of gas has on the HCN emission can be seen in Figure \ref{fig:NH2W} and we see that as the column density increases the intensity of HCN emission also increases. It appears that HCN is above the observationally detectable limits of $\sim 0.1\ \mathrm{K\ km\, s^{-1}}$ only above a column density of $\sim 3 \times 10^{21} \mathrm{cm^{-2}}$, which is roughly an order of magnitude higher than our lower threshold on the CO to HCN conversion in Figure \ref{fig:xhcnav}. A similar behaviour in the HCN emission is reported by \cite{PETY2017}, \cite{KAUFFMANN2017} and \cite{BARNES2020} who find $\sim 50\%$ emission stemming from column densities below $9.7 \times 10^{21}\ \mathrm{cm^{-2}}$, $1.2 \times 10^{22}\ \mathrm{cm^{-2}}$ and $2.1 \times 10^{22}\ \mathrm{cm^{-2}}$ respectively. 

The inclusion of sink particles in our figures allows us to clearly demonstrate that the regions of active star formation are associated with bright HCN emission. However, upon looking at cases C and F (see Table \ref{tab:tableoverview}), we also see sink particles without any HCN emission demonstrating that star forming clouds can evolve rapidly: stars can be ejected from their natal environments, and young clusters may also consume the available gas on the local free-fall time. Note that there is no feedback from the sink particles in our simulations. 

The main goal of this paper is to determine the density regime probed by HCN.   We achieve this by manipulating the HCN abundance that goes into RADMC-3D. By artificially setting the HCN abundance in cells with a density  below a certain threshold density, $\mathrm{n_{thr}}$, to zero in our RADMC-3D input cubes, and then performing the RT for the HCN line, we can determine the amount of HCN emission arising from gas with $n > \mathrm{n_{thr}}$.
If we repeat this process, systemically varying $\mathrm{n_{thr}}$  from $\mathrm{10^2\ to\ 10^6\ cm^{-3}}$, we can work out the inverse cumulative fraction of HCN emission with density for each cloud.
This analysis is repeated for all twelve of the simulated regions studied in this paper. However, it is standard to determine the cumulative fraction of emission i.e. the amount of HCN emission arising from gas with $n < \mathrm{n_{thr}}$. Therefore to do this we simply take one minus our inverse cumulative fraction of HCN.

The results of this process can be seen in Figure \ref{fig:fracemall}, in which the fraction of emission is given by the ratio of the HCN luminosity produced by gas below a certain density threshold, $\mathrm{n_{thr}}$, to that in the case where no threshold is applied -- i.e. the HCN abundance is unchanged from the value derived from Equation \ref{eq:xHCN}.  In Figure \ref{fig:fracemall}, we see $\sim\ 50\%$ of HCN emission emanates from densities below $\sim \mathrm{1-7 \times 10^3\ cm^{-3}}$, with the scatter depending on both the evolutionary stage of the cloud in the simulation, and the collision velocity (the former proving a slightly larger scatter). This result goes against many observational results that postulate that most of the emission stems from densities above $\mathrm{10^4\ cm^{-3}}$ such as \cite{GAO2004a, KRUMHOLZ2007}. However, our result agrees with the more recent observational studies such as \cite{ SHIRLEY2015, KAUFFMANN2017, PETY2017, HARADA2019, TAFALLA2021}. 

We see a trend towards a higher fraction of emission coming from higher densities as the simulations evolve over time, as they are able to accumulate a higher fraction of dense gas (as shown in Figure \ref{fig:evolmass}). Therefore, one could argue that we could reach a point where the emission from dense gas overwhelmingly dominates. However, we see both observationally and through simulations that the fraction of gas above $\mathrm{10^4 cm^{-3}}$ is generally small \citep{KAINULAINEN2009, LADA2010}, at least outside the galactic centre \citep{LONGMORE2013}.  It is therefore unclear if an environment with enough dense gas for the HCN emission to probe densities above  $\mathrm{3 \times 10^4\ cm^{-3}}$ is common in galaxies like the Milky Way, outside ``extreme'' environments. 

The idea that more dense gas equates to a higher threshold density for HCN emission actually breaks down as soon as we vary our cloud-cloud collision velocity. We see a trend of decreasing threshold density as we increase our initial cloud velocity, even though we generally see an increase in the mass of gas at higher densities as we increase the collision velocity (Figure \ref{fig:evolmass}). The reason behind this is simply that the fraction of mass residing at densities above $\mathrm{10^{3} \, cm^{-3}}$  decreases as the initial cloud velocity of the simulation increases (see Table \ref{tab:tableoverview}). As we will see below, it is around this density that is best traced by HCN (1-0) emission. 

We can compare our results to those of \cite{KAUFFMANN2017} by adopting their definition of the `characteristic density', $\mathrm{n_{char}}$, which is the density below which half of the total integrated intensity arises, i.e. $\mathrm{W_{HCN}\ (n\ <\ n_{thr})}\ /\ \mathrm{W_{HCN, Total}}  = 50 \%$.  From the data in Figure \ref{fig:fracemall}, we determine that $\mathrm{n_{char}}$ is $2.85${\raisebox{0.5ex}{\tiny$^{+4.25}_{-2.15}$}} $\times\ 10^3$ $\mathrm{cm^{-3}}$ for our suite of simulations (i.e. taking the mean from all our RT modelling).  Our characteristic density lies in between those derived for Orion A -- $0.87${\raisebox{0.5ex}{\tiny$^{+1.24}_{-0.55}$}} $\times\ 10^3$ $\mathrm{cm^{-3}}$  \citep{KAUFFMANN2017} -- and W49 -- $3.4\pm 2.8\times 10^3$ $\mathrm{cm^{-3}}$ \citep{BARNES2020} . 

A further comparison between our work and both \cite{KAUFFMANN2017} and \cite{BARNES2020}, can be made by looking at the cumulative fraction of emission as a function of $\mathrm{A_V}$. Our column densities are derived by integrating along the $z$-direction in the RADMC-3D density cubes, to ensure that it is consistent with the 3D structure used in the RT (as opposed to deriving it straight from {\sc Arepo}'s more detailed Voronoi grid).  The resulting column densities are then converted to $\mathrm{A_V}$ using Equation \ref{eq:AV}. The cumulative total emission as a function of $\mathrm{A_V}$ is given in Figure \ref{fig:fracemav}. We find that our results are closer to those found for Orion A by \citet{KAUFFMANN2017}, than the results from W49 by \citet{BARNES2020}. We discuss this further in Section \ref{sec:disc}. 

\begin{table}
	\centering
	\caption{Summary of our findings of both the characteristic density and characteristic visual extinction traced by HCN emission. For comparison, we also quote the values from two recent observational studies.}
	\label{tab:tablesummary}
	\begin{tabular}{lcccr} 
		\hline
		& Reference & $\mathrm{A_{V,char}}$ & $\mathrm{n_{char}}$ & \\
		& & [$\mathrm{mag}$] & [$\times 10^3$ $\mathrm{cm^{-3}}$] & \\
		\hline
		& Jones et al. [this paper] & $5.05${\raisebox{0.5ex}{\tiny$^{+3.36}_{-3.07}$}} & $2.85${\raisebox{0.5ex}{\tiny$^{+4.25}_{-2.15}$}} & \\
		& \cite{KAUFFMANN2017} & $6.1${\raisebox{0.5ex}{\tiny$^{+1.2}_{-1.0}$}} & $0.87${\raisebox{0.5ex}{\tiny$^{+1.24}_{-0.55}$}} & \\
		& \cite{BARNES2020} & $11.9\pm 1.1$ & $3.4\pm 2.8$ & \\
		\hline
	\end{tabular}
\end{table}

We can perform a similar analysis for the characteristic visual extinction, and define $\mathrm{A_{V,char}}$, where $\mathrm{A_V}$ contains half the total integrated intensity, or $\mathrm{W_{HCN}\ (A_V\ <\ A_{V, thr})}\ /\ \mathrm{W_{HCN, Total}} = 50 \%$.  From  the data used to compile Figure \ref{fig:fracemav}, we determine $\mathrm{A_{V,char}}$ to be $5.05${\raisebox{0.5ex}{\tiny$^{+3.36}_{-3.07}$}} $\mathrm{mag}$. 

While our value of $\mathrm{A_{V,char}}$ is consistent with that from Orion A, we can see from Figure \ref{fig:fracemav} that our simulations are not consistent with the data from W49. The fact that we see a different relation to \citet{BARNES2020} is not that surprising. First, their resolution is much cruder, nearly $\mathrm{\sim 3\ pc}$, yet they report high gas column densities. Given the low resolution, their high column density regions are likely probing much higher densities than in the Orion A observations, to compensate for the low densities that are likely mixed into emission within the beam. Second, W49 is a much denser region than those we study here, and so likely contains more gas at high densities than our clouds. Another big difference between W49 and the clouds modelled here is cloud mass. Our total available mass is $\mathrm{2 \times 10^4\ M_{\odot}}$, of which $\mathrm{\sim 10 - 20\, \%}$ is in the region we study. On the other hand, the mass of the region \citet{BARNES2020} map in W49 is $\mathrm{\sim 2 \times 10^5\ M_{\odot}}$, and the mass of the entire W49 complex is larger still, $\mathrm{\sim 10^6\ M_{\odot}}$.

Finally, the observations of W49A focused around the star-forming region of W49A and not the entire W49 region. As noted by \citet{PETY2017}, HCN is sensitive to far-UV radiation that is produced from star formation. Our ISRF with G$_0 = 1.7$ is hence not representative of the radiation field found in the massive star forming region of W49A. In contrast we see a good comparison with the data from \citet{KAUFFMANN2017}, where the spatial resolution is very similar to that in our study: roughly $\sim 0.02\ \mathrm{pc}$ in this work, compared to $\sim 0.05\ \mathrm{pc}$ in the case of the Orion A observations.  Along with \citet{KAUFFMANN2017} and \citet{BARNES2020}, we see that our data is also consistent to that of \citet{PETY2017}.

Finally, we can follow the analysis in \citet{KAUFFMANN2017} by examining the emission efficiency ratio, $\mathrm{h_{HCN}\ =\ W_{HCN}/N_{H_2}}$. In Figure \ref{fig:h}, we see the normalized ratio of the integrated intensity to the column density [or $\mathrm{h_{HCN}/h^{max}_{HCN}}$] as a function of column density. We see a clear trend with all of our simulations in that they all peak at around a column density of $\mathrm{10^{22}\ cm^{-2}}$ which is roughly equivalent to $\mathrm{A_V}$ = 5 mag which is again comparable to our findings above.



\begin{figure}
    \includegraphics[width=80mm]{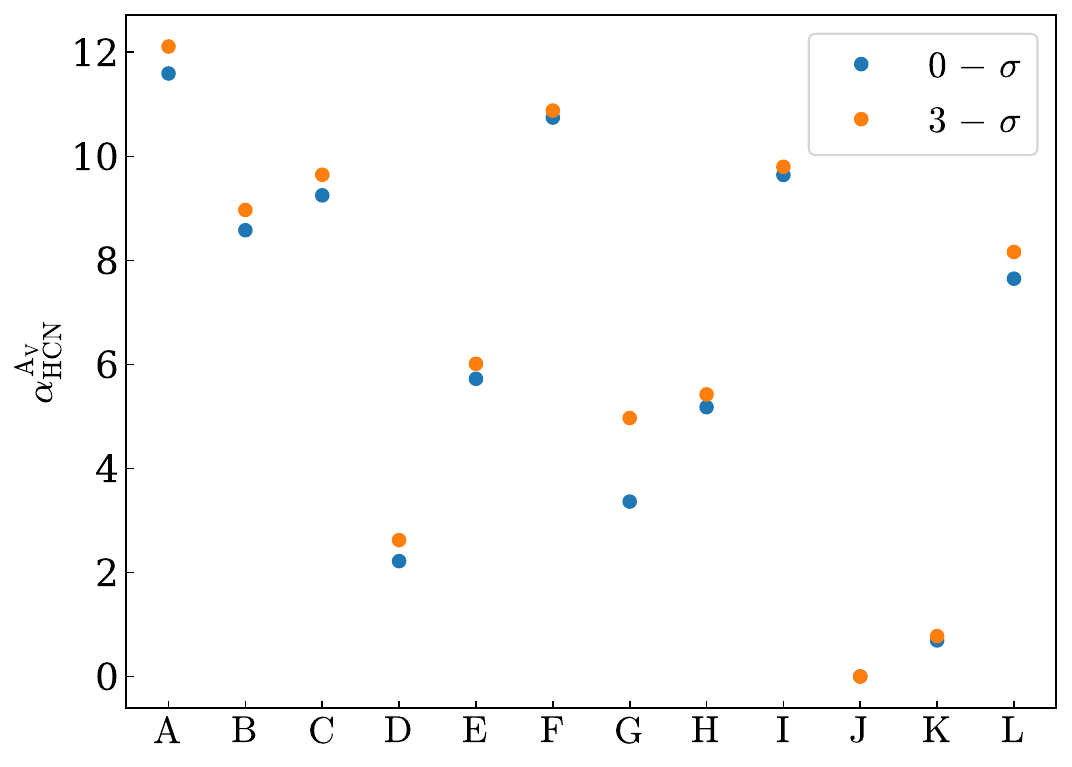}
    \caption{The conversion factor, $\mathrm{ \alpha_{HCN}^{A_V} }$, in each of the twelve cases considered in this paper. Note that the conversion factors here correspond to the use of an $\mathrm{A_V}$ above 8 mag as seen in Equation \ref{eq:alpha8av}. Here, the ID's of each simulation corresponding to Table \ref{tab:tableoverview} are given. The orange points show the values we obtain if we restrict the calculation to pixels where HCN is detected with signal-to-noise > 3 for an assumed noise level $\mathrm{\sigma \ = 0.1\ (K\, km\, s^{-1})}$. The blue points show the result in the ideal noise-free case. We see that in most cases, the inclusion of a realistic amount of noise makes very little difference to the derived value of $\mathrm{ \alpha_{HCN}^{A_V} }$.}    
    \label{fig:alpha}
\end{figure}

\begin{figure}
    \includegraphics[width=80mm]{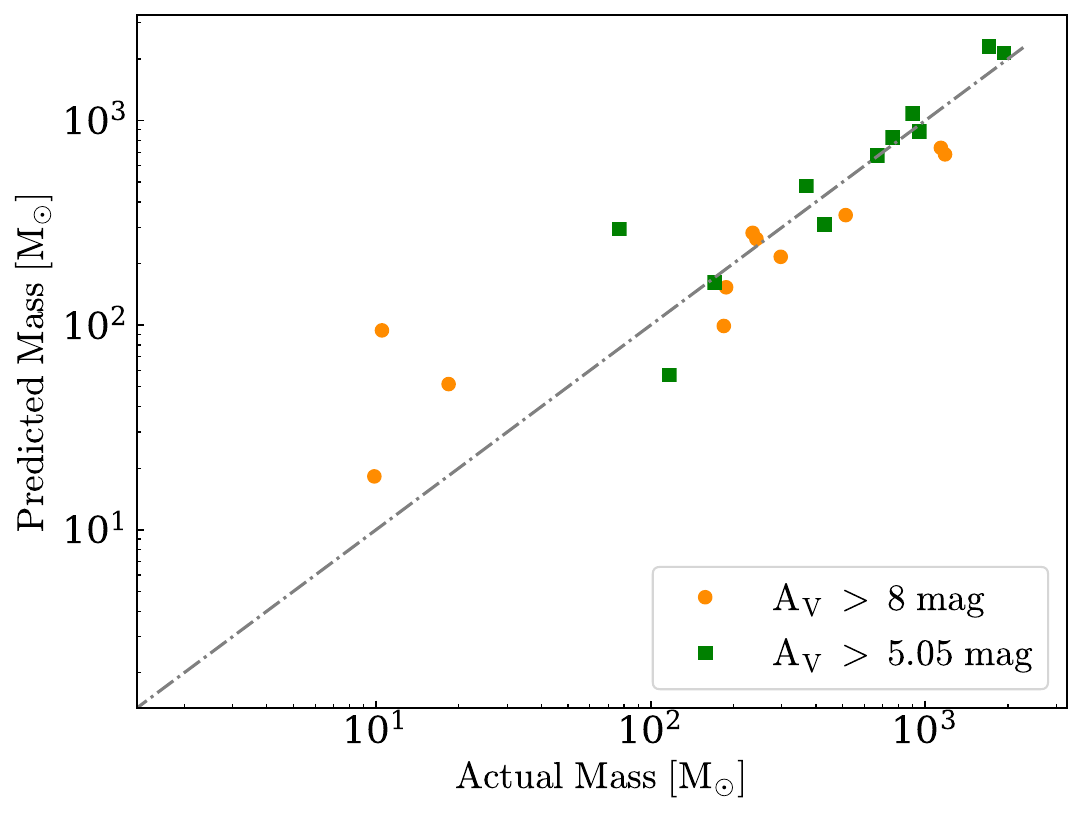}
    \caption{A plot of predicted mass for all simulations using a re-arranged version of equation \ref{eq:alpha8av} where $\mathrm{\alpha_{HCN}\ =\ 6.79\ M_{\odot}\, (K\, km\, s^{-1}\, pc^2)^{-1}}$ for $\mathrm{A_V\ >\ 8\ mag}$ compared to the actual mass calculated within regions of $\mathrm{A_V\ >\ 8\ mag}$ and $\mathrm{\alpha_{HCN}\ =\ 19.46\ M_{\odot}\, (K\, km\, s^{-1}\, pc^2)^{-1}}$ for $\mathrm{A_V\ >\ 5.05\ mag}$ compared to the actual mass calculated within regions of $\mathrm{A_V\ >\ 5.05\ mag}$. The dashed line denotes the point at which the predicted mass is equivalent to the actual mass.}
    \label{fig:massav}
\end{figure}

\begin{figure}
    \includegraphics[width=80mm]{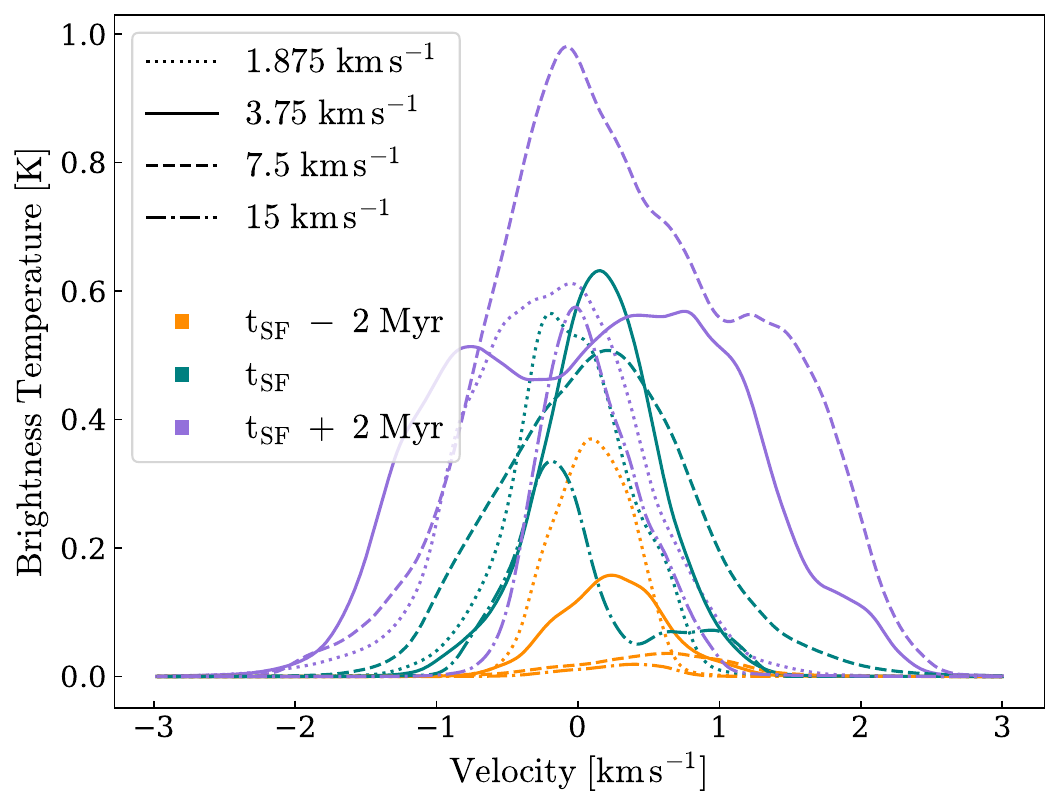}
    \caption{ HCN (1-0) spectra of all simulations. The lines styles and colours are the same as Figure \ref{fig:cdf}. We take the mean emission of our entire 10pc box. This is repeated for all velocity channels for all simulations to produce the observed spectra. }    
    \label{fig:lineplot}
\end{figure}

\begin{figure}
    \includegraphics[width=80mm]{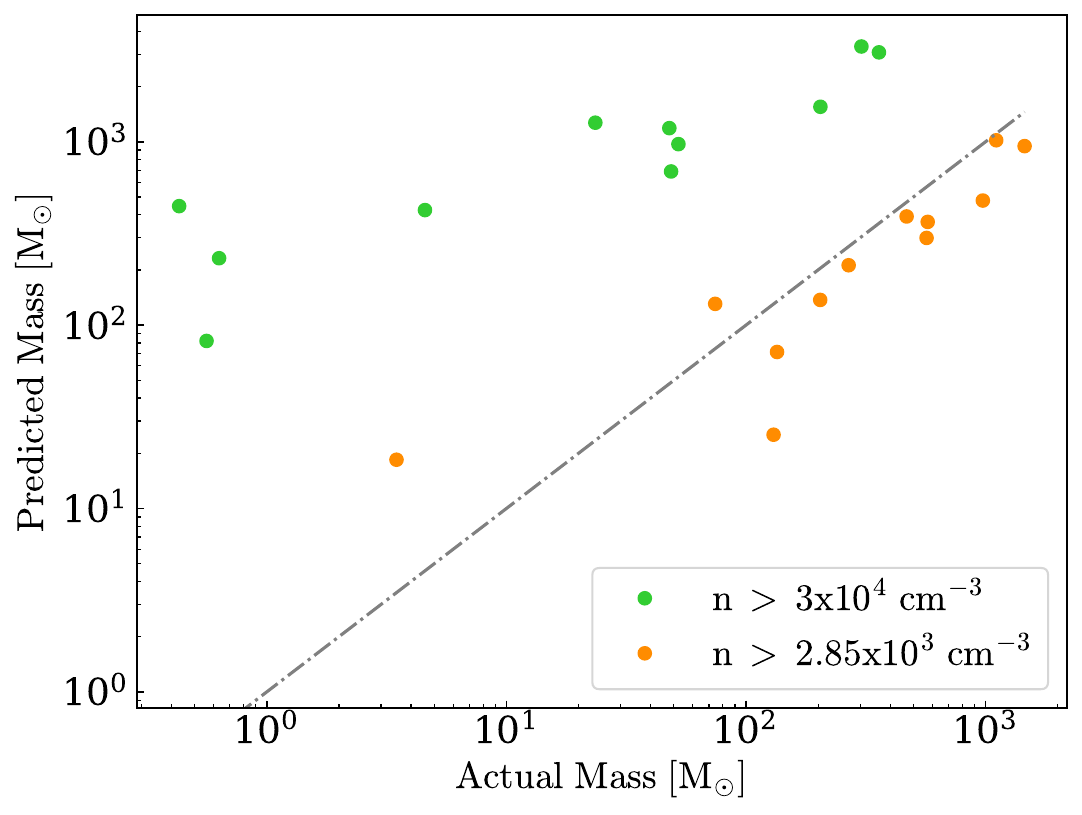}
    \caption{A plot of predicted mass for all simulations using Equation \ref{eq:massgaosol} for our density threshold of $\mathrm{2.85 \times 10^{3}\ cm^{-3}}$ and \citet{GAO2004a} density threshold of $\mathrm{3 \times 10^{4}\ cm^{-3}}$ compared to the actual mass calculated above these two density thresholds. The dashed line denotes the point at which the predicted mass is equivalent to the actual mass.}    
    \label{fig:massdens}
\end{figure}

\subsection{Using HCN emission to determine the mass of `dense' gas}
\label{subsubsec:HCNalpha}

HCN emission has been used as way of tracing `dense' gas in molecular clouds, with a conversion factor of the form, 
\begin{equation}
\mathrm{\alpha_{HCN}\ =\ M_{dg}/L_{HCN}},
\label{eq:alphadg}
\end{equation}
where $\mathrm{\alpha_{HCN}}$ is units of $\mathrm{M_{\odot}\, (K\, km\, s^{-1}\, pc^2)^{-1}}$.  The typically adopted value is $\mathrm{\alpha_{HCN}\ \sim\ 10 M_{\odot}\, (K\, km\, s^{-1}\, pc^2)^{-1}}$ \citep{GAO2004b}.  As already discussed, there is some uncertainty as to what exactly `dense' means here, with (somewhat confusingly) both definitions based on volume and column density being used in the literature, as well as uncertainty over the value of the density being used in each case. In this section we explore both the column density ($\alpha_{\rm HCN}^{\rm A_V}$) and volume density ($\alpha_{\rm HCN}^{n_{\rm thr}}$) versions of this conversion factor. 

We start our discussion by looking at the conversion factors based on column density, or in this case visual extinction, such as explored by \citet{KAUFFMANN2017, EVANS2020, BARNES2020}. For clarity we will define, 
\begin{equation}
\mathrm{\alpha_{HCN}^{A_V}\ =\ M_{A_V\, >\, 8\, mag}/L_{HCN}}
\label{eq:alpha8av}
\end{equation}
where $\mathrm{M_{A_V\, >\, 8\, mag}}$ is the mass residing above an $\mathrm{A_V > 8 mag}$, as given in \citet{EVANS2020} and \citet{BARNES2020}. This is in keeping with the result from  \cite{LADA2010, LADA2012}, who defined a ``threshold'' surface density of $\mathrm{8\ mag}$ above which the vast majority of dense cores are found.  Note however that \cite{KAUFFMANN2017} uses a visual extinction of $\mathrm{7\ mag}$ in their analysis. 

In Figure \ref{fig:alpha} we present the conversion factor, $\mathrm{ \alpha_{HCN}^{A_V} }$, as derived from our simulation data, adopting the definition given in Equation \ref{eq:alpha8av}. To mimic the effects of limited observational sensitivity, we only consider pixels of integrated emission -- and thus the corresponding pixels in the column density ($\mathrm{A_V}$) maps -- that would be detected with a signal-to-noise greater than 3 for an assumed uniform noise level of $\mathrm{\sigma \ = 0.1\ (K\, km\, s^{-1})}$. For comparison, we also show the results in the noise-free case. 

It is clear from Figure \ref{fig:alpha} that there is a broad scatter in the values of $\mathrm{ \alpha_{HCN}^{A_V} }$ derived from our simulations, and that the values are largely unaffected by the choice of $\mathrm{\sigma}$. We therefore ignore the effect of $\mathrm{\sigma}$ on our analysis. The mean value of $\mathrm{ \alpha_{HCN}^{A_V} }$ from our suite of simulations is $\mathrm{\alpha_{HCN}^{A_V}\ =\ 6.79\ M_{\odot}\, (K\, km\, s^{-1}\, pc^2)^{-1}}$.  Although this is close to the value quoted in \citet{GAO2004b}, the standard deviation is $\mathrm{3.79\ M_{\odot}\, (K\, km\, s^{-1}\, pc^2)^{-1}}$ which is over half the mean value of $\mathrm{\alpha_{HCN}^{A_V}}$.

We can demonstrate the effect of adopting an incorrect $\mathrm{\alpha_{HCN}^{A_V}}$ by determining the mass we would expect to get above an $\mathrm{A_V}$ of $\mathrm{8\ mag}$ using Equation \ref{eq:alpha8av} and comparing to the actual mass in our simulations that lies above an $\mathrm{A_V}$ of $\mathrm{8\ mag}$. We show the results of this analysis in Figure \ref{fig:massav}. We see that for a low amount of dense gas -- around 10 M$_{\odot}$, such as one might find in a handful of prestellar cores -- the standard $\mathrm{\alpha_{HCN}}$ relation overestimates the amount of dense gas present by up to an order of magnitude. However, once the clouds contain $\mathrm{M_{A_V\, >\, 8\, mag} > 100 M_{\odot}}$,  the standard $\mathrm{\alpha_{HCN}^{A_V}}$ actually predicts the amount of dense gas very well.  This suggests that provided one is already looking at well-evolved, and active star-forming regions, the true scatter in $\mathrm{\alpha_{HCN}^{A_V}}$ will not significantly affect the predicted mass.  However, for regions of low star formation (i.e. early in a cloud's star-forming evolution), one could significantly over-predict the amount of dense gas present.

We can also do the same analysis using our value of $\mathrm{A_{V, char }\ =\ 5.05\ mag}$ that we we derived above, instead of $\mathrm{A_V}$ of $\mathrm{8\ mag}$ in Equation \ref{eq:alpha8av}. Here, we get a mean value of $\mathrm{\alpha_{HCN}^{A_V} \ =\ 19.46\ M_{\odot}\, (K\, km\, s^{-1}\, pc^2)^{-1}}$. We can use this value of $\mathrm{\alpha_{HCN}^{A_V} }$ to repeat the analysis we just performed with $\mathrm{A_V}$ of $\mathrm{8\ mag}$ for our $\mathrm{A_{V, char}}$ of $\mathrm{5.05\ mag}$, which can be seen in Figure \ref{fig:massav}. We see that using an $\mathrm{A_{V, char}}$ of $\mathrm{5.05\ mag}$ is a much more reliable estimate of mass across all stages of our suite of simulations.

We can now perform a similar analysis for the conversion factor between HCN emission and gas above a threshold volume density,  $\mathrm{\alpha_{HCN}^{n_{thr}}}$. As described in \citet[][see also \citealt{BARNES2020}]{GAO2004a}, the mass above the threshold density can be given by,
\begin{equation}
\mathrm{M_{n\, >\, n_{thr}}^{sum}\ \approx\ 2.1 \dfrac{n_{thr}^{0.5}}{T_{B}}  L_{\rm HCN}},
\label{eq:massgaosol}
\end{equation}
where $\mathrm{T_{B}}$ is the intrinsic HCN line brightness temperatures and $n_{\rm thr}$ is our density threshold. The factor of $2.1 \rm n_{\rm thr}^{0.5} / {\rm T}_{B}$ can hence be identified with the volume density based conversion factor, $\alpha_{\rm HCN}^{n_{\rm thr}}$. 

The intrinsic brightness temperature for Equation \ref{eq:massgaosol} would simply be the peak main beam temperature of the spectra, provided that we are dealing with extended sources (i.e. the source is filling the beam) like \citealt{BARNES2020}. However, if we take our box to be the beam we clearly have a case where the source is much smaller than the telescope beam. Indeed, we see this in Figure \ref{fig:lineplot} where the main beam temperatures are at least an order of magnitude lower than those seen in \citealt{BARNES2020}. In this case we would have to take into account the solid angle of the source convolved with the diameter (FWHM) of the telescope Gaussian beam $\mathrm{\theta_{MB}}$ (see section 4.2 in \citealt{GAO2004a}). Since all the observational studies have a different set-up, we instead use our pixel size as the main beam size and therefore our intrinsic brightness temperature is simply our pixel brightness temperature.

From our simulations, we find $\mathrm{T_{B}}$ to vary between 6 and 14 K, with a mean of roughly $\mathrm{13\ K}$. We now consider two values for ${n_{\rm thr}}$: first, we input the density that is most commonly adopted in the literature \citep{GAO2004a} of $\mathrm{3 \times 10^4\ cm^{-3}}$; and second we input our characteristic density of $\mathrm{2.85 \times 10^3\ cm^{-3}}$ that we established from our simulations in Section \ref{subsubsec:HCNdensity}.  This yields a conversion factor of $\alpha_{\rm HCN}^{n_{\rm thr}} = 8.62\ \rm M_{\odot}\, (K\, km\, s^{-1}\, pc^2)^{-1}$ for $ n\ >\ 2.85 \times 10^3\ \rm cm^{-3}$ and $\alpha_{\rm HCN}^{n_{\rm thr}} \ =\ 27.98\ \rm M_{\odot}\, (K\, km\, s^{-1}\, pc^2)^{-1}$ for $ n\ >\ 3 \times 10^4\ \rm cm^{-3}$.

We see from Figure \ref{fig:massdens} that our lower calculated density of $\mathrm{2.85 \times 10^3\ cm^{-3}}$ is able to reproduce the actual mass far better than $\mathrm{3 \times 10^4\ cm^{-3}}$. In contrast, the standard value of  $n_{\rm thr} = \mathrm{3 \times 10^4\ cm^{-3}}$ consistently over-predicts the amount of mass above this density in our simulations by at least an order of magnitude; in one extreme case  it predicts $\mathrm{59.9\ M_{\odot}}$ above $\mathrm{3 \times 10^4\ cm^{-3}}$  even though {\em the simulation contained no mass above this density}.

\subsection{The effect of optical depth on the HCN emission}
\label{subsubsec:relateHCNtau}


We can take advantage of the fact that RADMC-3D can produce optical depth maps of our simulations to explore the effect that the optical depth has on the emission of HCN. Selecting this option in RADMC-3D generates a PPV cube of optical depths (we will use $\mathrm{\tau_{HCN}}$ for the remainder of the paper) instead of emission. We can then use these $\mathrm{\tau_{HCN}}$ PPV cubes to find out if the use of $\mathrm{n_H \gtrsim 6 \times 10^{4} / \tau \ cm^{-3}}$ by the likes of \citealt{GAO2004a} and \citealt{KRUMHOLZ2007} is justifiable since we can use our $\mathrm{\tau_{HCN}}$ to predict an effective density and compare it to our effective density from Table \ref{tab:tablesummary}. Note that we restrict our analysis to the $\mathrm{t_{SF}\ +\ 2\ Myr}$ snapshots as this is where we find the most dense gas (e.g. Fig \ref{fig:cdf}).

\begin{figure*}
    \centerline{
    \includegraphics[width=81mm]{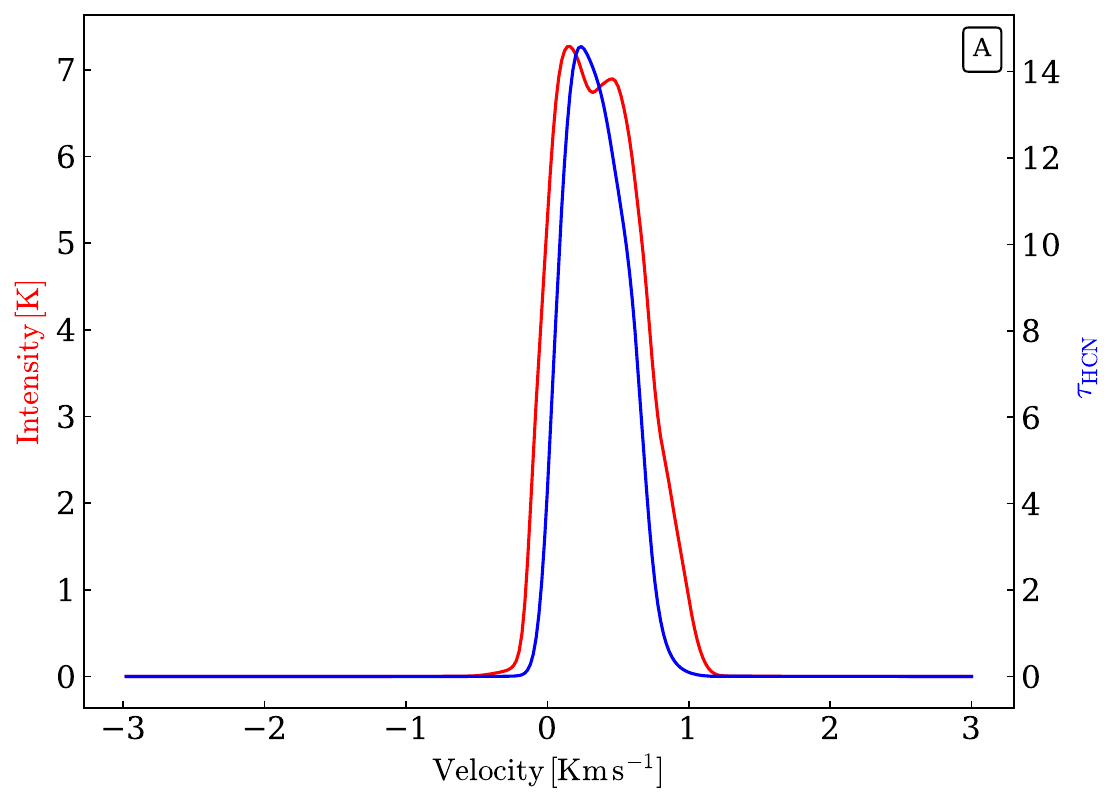}
    \includegraphics[width=81mm]{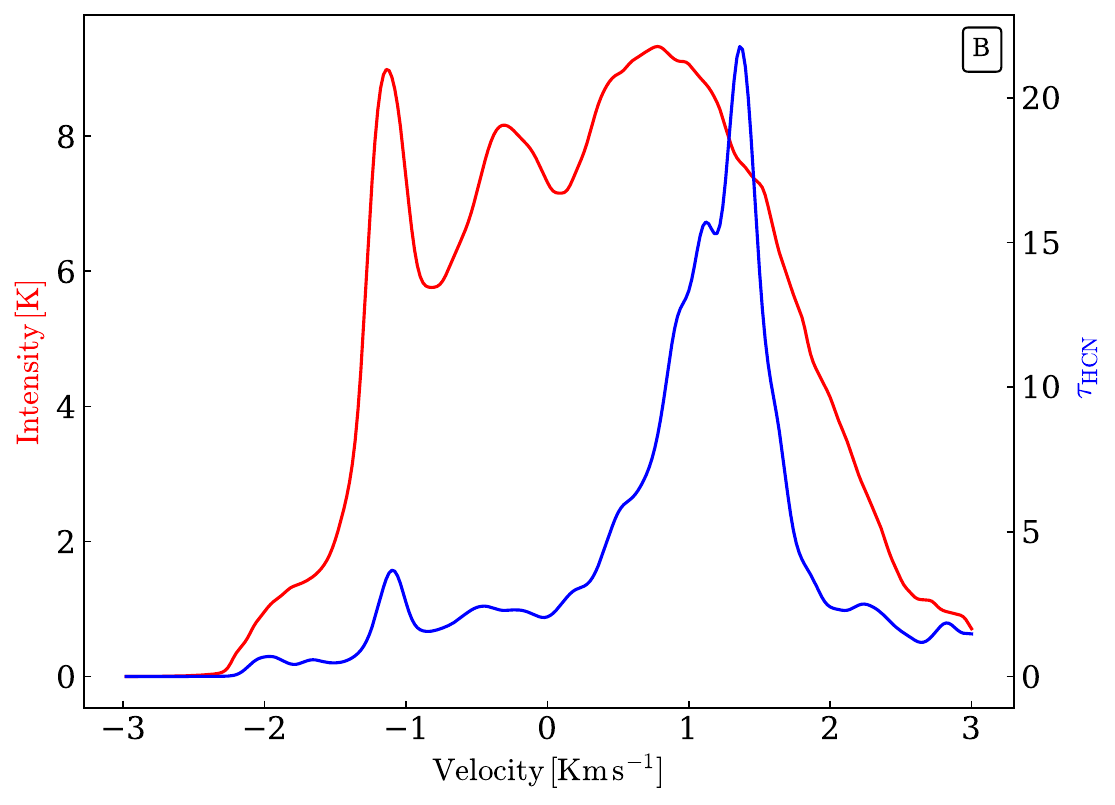}
    }
    \vspace{0mm}
    \centerline{
    \includegraphics[trim={0 0 0 4.25cm}, clip, width=180mm]{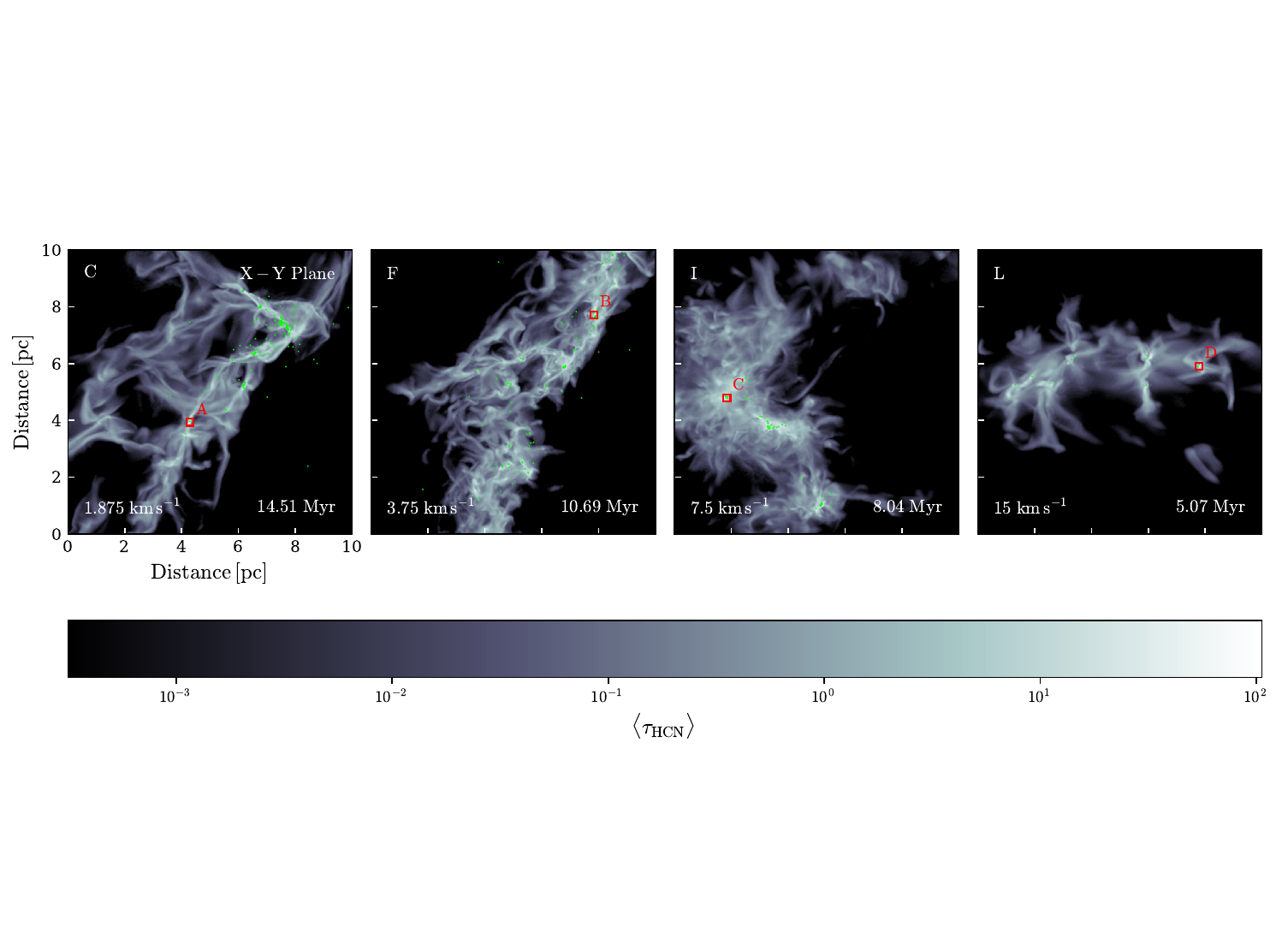}}
    \vspace{-25mm}
    \centerline{
    \includegraphics[width=81mm]{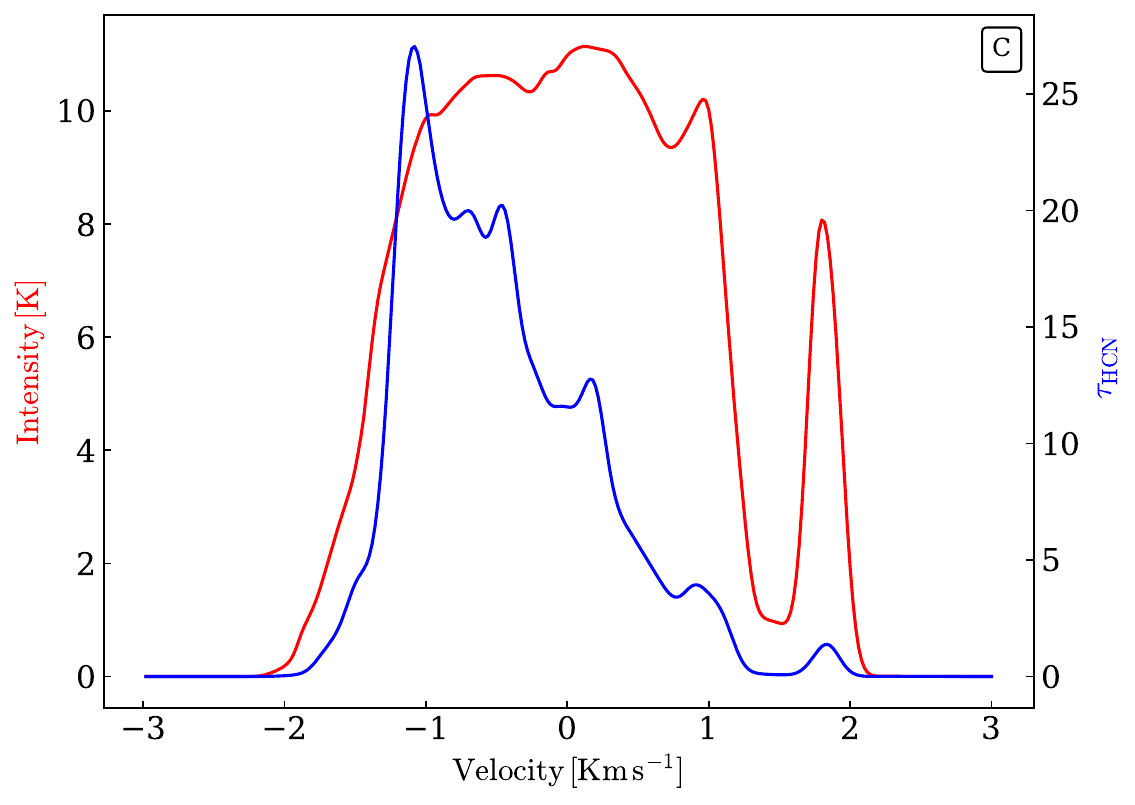}
    \includegraphics[width=81mm]{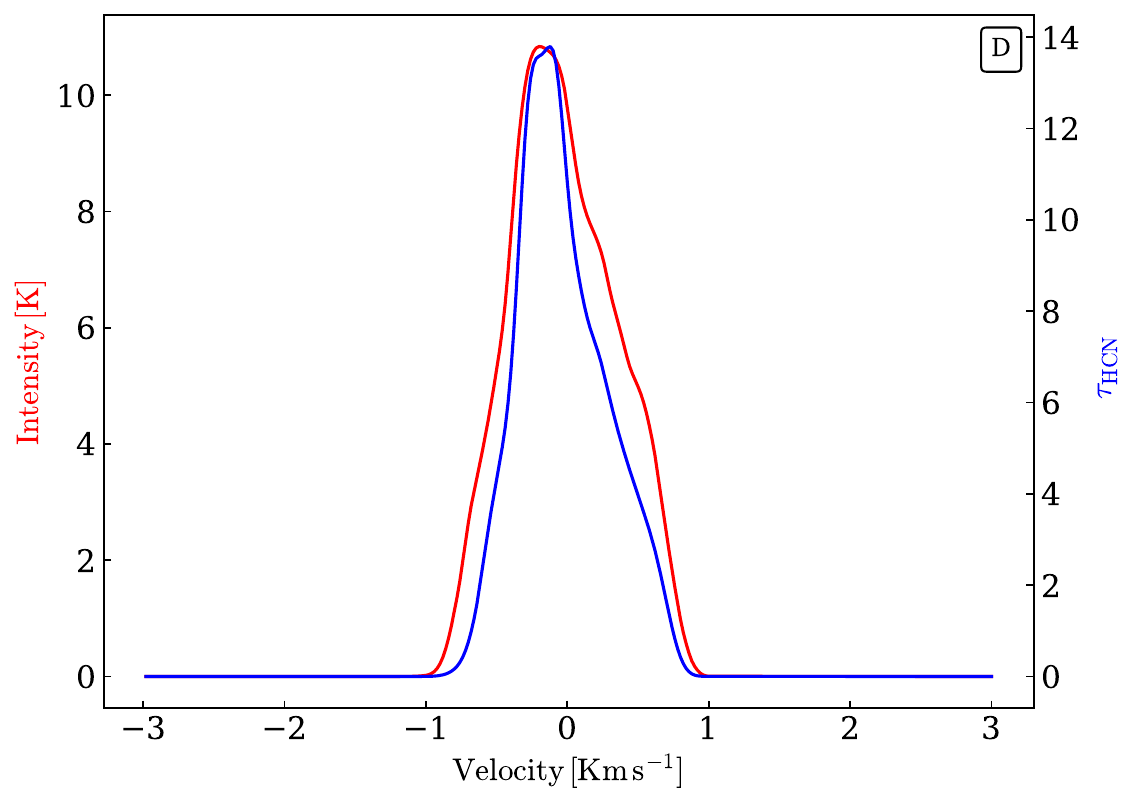}
    }    
    \caption{Emission-weighted integrated optical depth maps of the four initial velocities, all at the later simulation time surrounded by four line spectra and optical depths of selected regions labelled alphabetically (in red) on the optical depth maps. These regions are chosen with the peak optical depth as the centre of a square region with a size of 10 pixels, taking the mean emission and the mean optical depth of these square regions and repeating through each velocity channel to produce the resulting four line profile plots.}    
    \label{fig:tauplot}
\end{figure*}

We can create maps of the mean optical depth along a line of sight by defining,
\begin{equation}
\mathrm{ \langle \tau_{HCN} \rangle = \frac{\sum_{i} \tau_{i} \times T_{i}}{\sum_{i} T_{i}} },
\label{eq:tauHCN}
\end{equation}
where i denotes the index of the PPV cube along the velocity axis. The  maps of mean optical depth for all four initial velocities are shown in Figure \ref{fig:tauplot}. 

\begin{figure*}
    \centerline{
    \includegraphics[width=81mm]{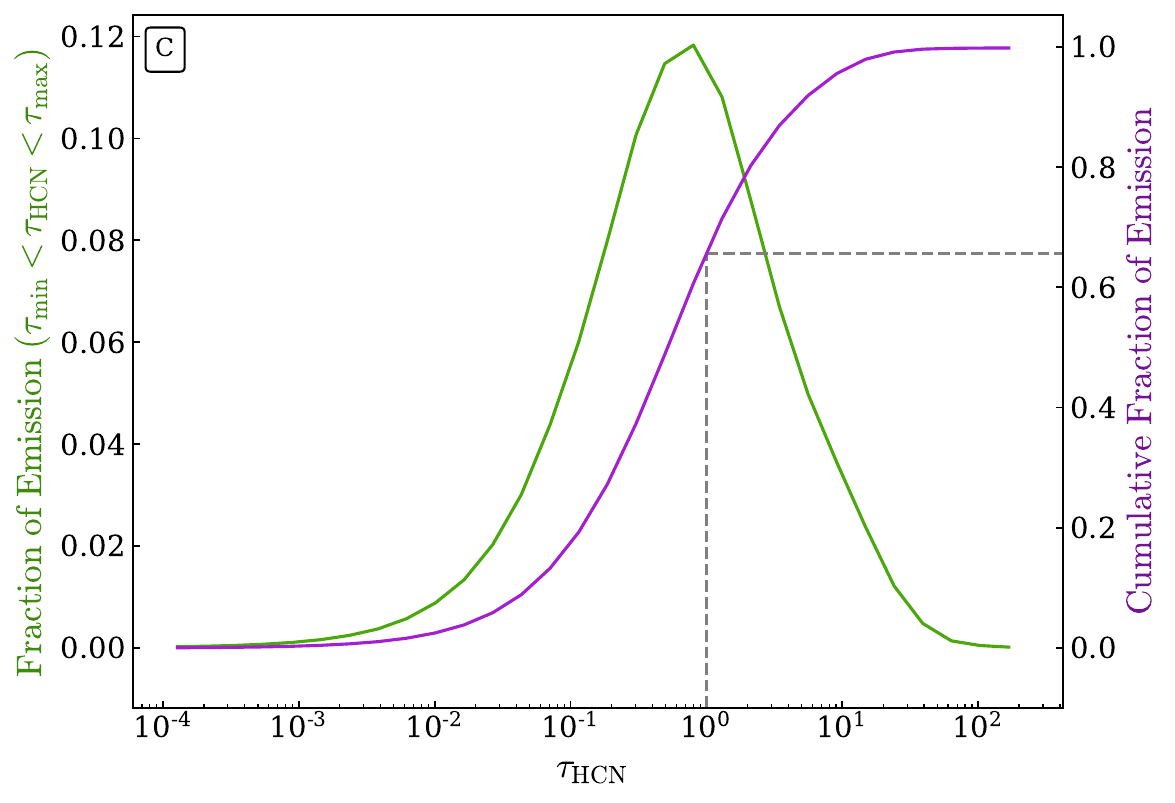}
    \includegraphics[width=81mm]{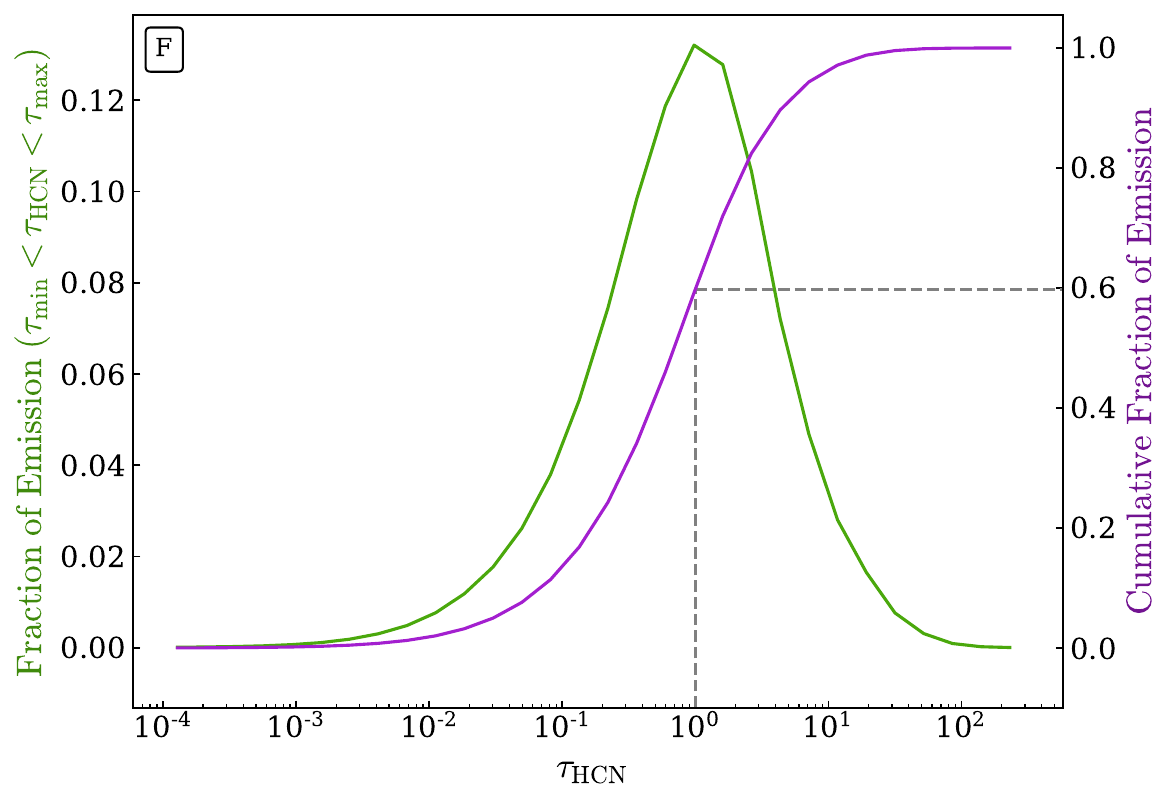}
    }
    \centerline{
    \includegraphics[width=81mm]{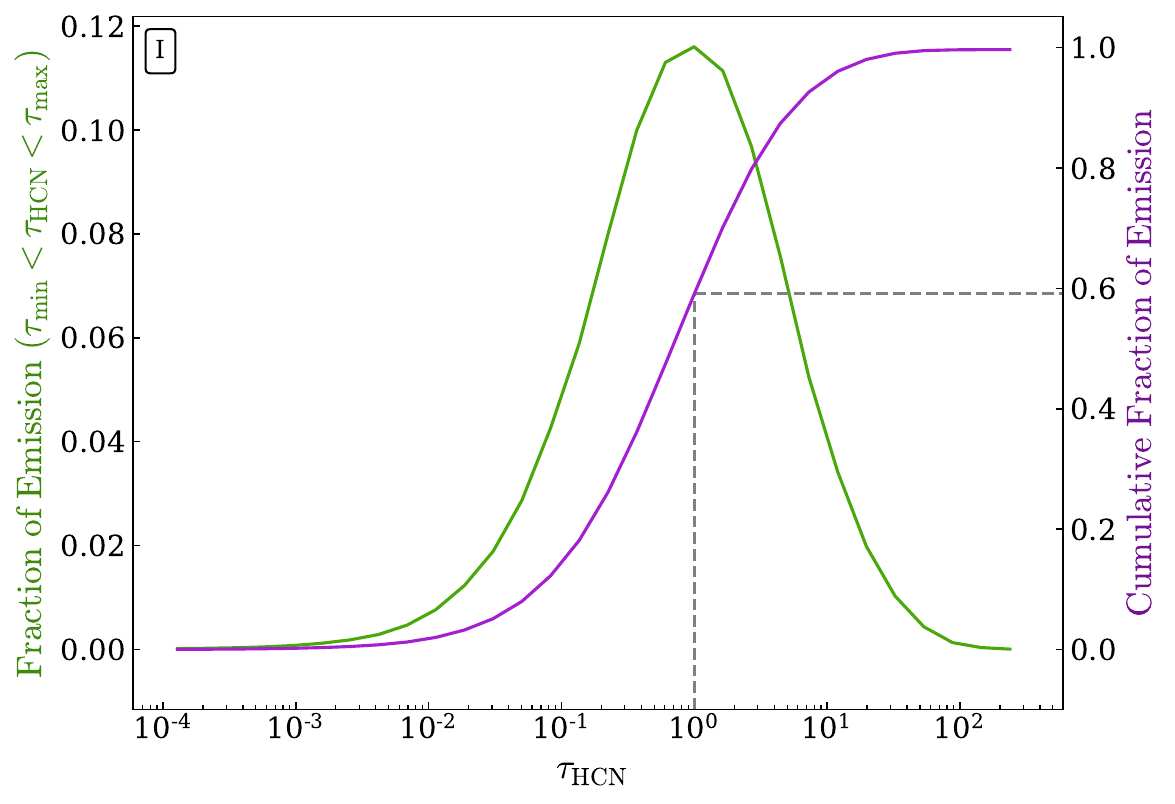}
    \includegraphics[width=81mm]{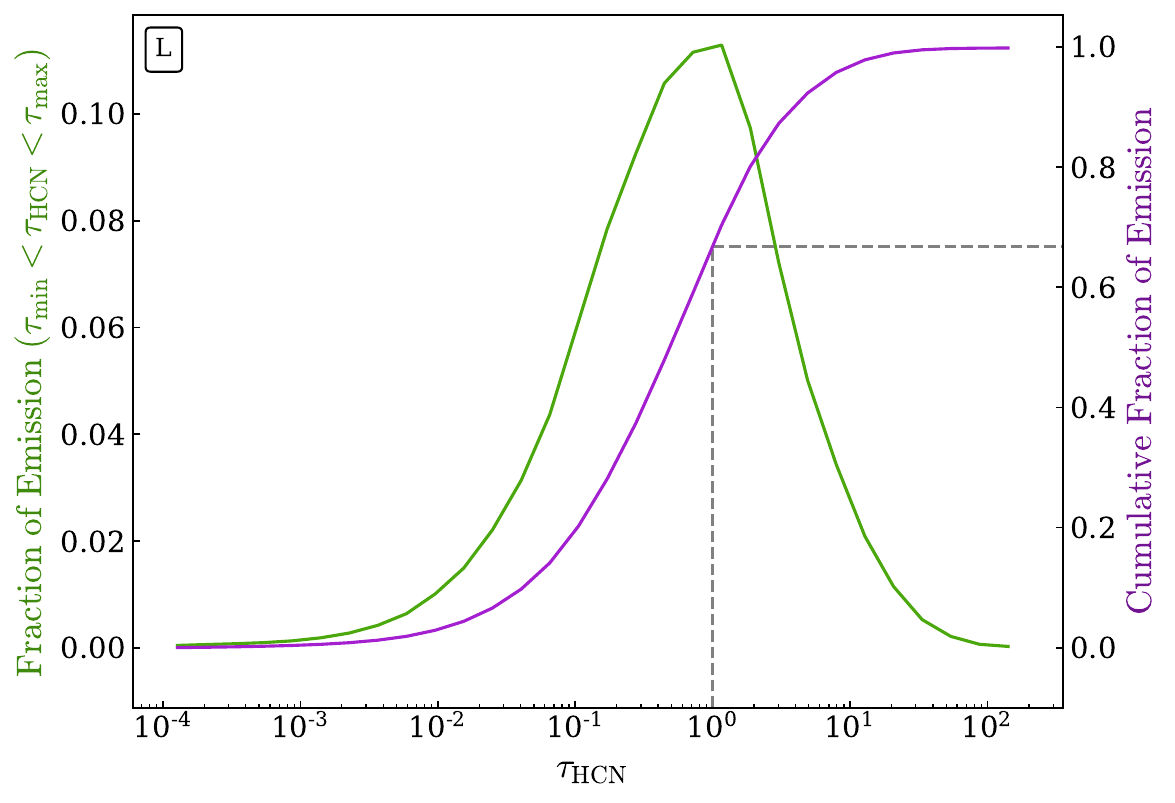}
    }    
    \caption{The fraction of HCN emission as function of optical depth (with accompanying cumulative fraction) for the four later stage simulations. The simulation IDs corresponds to those used in Table \ref{tab:tableoverview}. }    
    \label{fig:fractauplot}
\end{figure*}

We see from Figure \ref{fig:tauplot} that $\mathrm{\langle \tau_{HCN} \rangle}$ is high, above 10, towards the bright regions of HCN emission that we see in Figure \ref{fig:ivmap}, which are associated with column densities in excess of $10^{22} \rm cm^{-2}$ (see Figure \ref{fig:gridcol}). However $\mathrm{\langle \tau_{HCN} \rangle}$ is substantially lower than this for most of the map. To get a better idea of the optical depths associated with the bulk of the emission, we show in Figure \ref{fig:fractauplot} the fraction of emission of HCN as a function of $\mathrm{\tau_{HCN}}$. We see that the the peak fraction of emission occurs at around $\mathrm{\tau_{HCN} = 1}$ for all 4 simulations.  However, from the cumulative distributions, we see that only between 33 \% and 41 \% of the emission emanates from $\mathrm{\tau_{HCN} > 1}$, as summarised in Table \ref{tab:tablecumulativetau}.

\begin{table}
	\centering
	\caption{A table of the percentage of cumulative fraction of emission emanating from $\mathrm{\tau_{HCN}\ >\ 1}$ from Figure \ref{fig:fractauplot}.}
	\label{tab:tablecumulativetau}
	\begin{tabular}{lcccr} 
		\hline
		& ID & Percentage & \\
		& & [$\%$] & \\
		\hline
        & C & 34.5 & \\
        & F & 40.2 & \\
        & I & 40.9 & \\
        & L & 33.2 & \\
		\hline
	\end{tabular}
\end{table}

To give some indication of the variations of the optical depth with velocity (and thus along the line of sight), we also select a small region of $10\times10$ pixels within these simulations to compare the mean line spectra to the mean $\mathrm{\tau_{HCN}}$ line. These regions are focused on bright spots in the integrated HCN intensity, and are labelled alphabetically in Figure \ref{fig:tauplot}. In contrast to the low optical depth seen for the bulk of the cloud, the majority of the emission in these  $10\times10$ pixel regions stems from $\mathrm{\tau_{HCN} > 1}$; indeed, we find some 83 - 95 percent of the emission is associated with $\tau_{\rm HCN} > 1$. 

Our analysis shows that while the effective critical density of the HCN might be lowered towards bright, dense (and possibly pre-star-forming) cores, this is not the case for the bulk of the emission in a molecular cloud such as those we study here. We conclude that the low effective density for HCN in our study is the result of sub-thermal excitation from a large amount of low-density material, rather than a lowering of the critical density, such as suggested by \citet{BARNES2020} and \citet{GAO2004a}.

%
%
%

%
%
%

\section{The relationship between HCN/CO and dense gas.}
\label{subsubsec:relateHCNgas}

As well as using the brightness of the HCN (1-0) line, we can also use the ratio of the HCN (1-0) and CO (1-0) lines to constrain the distribution of gas volume densities in molecular clouds (e.g., \citealt{GAO2004b}, \citealt{GARCIA-BURILLO2012}, \citealt{LEROY2017}, \citealt{GALLAGHER2018}).
Because each line traces densities above their effective densities (for a clear definition of effective densities, see \citealt{SHIRLEY2015}), $\mathrm{n_{eff}}$, a change in the ratio of intensities between two lines with different $\mathrm{n_{eff}}$ can gauge changes in the estimated dense gas mass fraction (see, for example,  \citealt{KRUMHOLZ2007a} and \citealt{LEROY2017}). This multiple line method improves the accuracy with which variations in the sub-beam density distribution are recovered, and so is well-suited for low resolution studies i.e. galactic-scale studies which use high effective critical density lines that tend not to fill the beam.

Due to the high effective critical density of HCN, galactic scale studies have begun using HCN (1-0) to CO (1-0) integrated intensity ratio to estimate the distribution of gas volume density (e.g., \citealt{LEROY2017}, \citealt{GALLAGHER2018}, \citealt{QUEREJETA2019}). We therefore investigate how the HCN (1-0) / CO (1-0) integrated intensity ratio varies as a function of CO (2-1) line emission which is used as a proxy for surface density. We use the same RT code, RADMC-3D, that we use for our HCN analysis. For both CO (1-0) and (2-1) lines we use ortho- and para- $\mathrm{H_2}$ as collisional partners along with the collisional rates provided by Leiden Atomic and Molecular Database \citep{SCHOIER2005, JANKOWSKI2005, YANG2010}. When calculating the level populations of the CO, we assume that the gas has an H$_2$ ortho-to-para ratio of 3:1.

\begin{figure}
    \includegraphics[width=80mm]{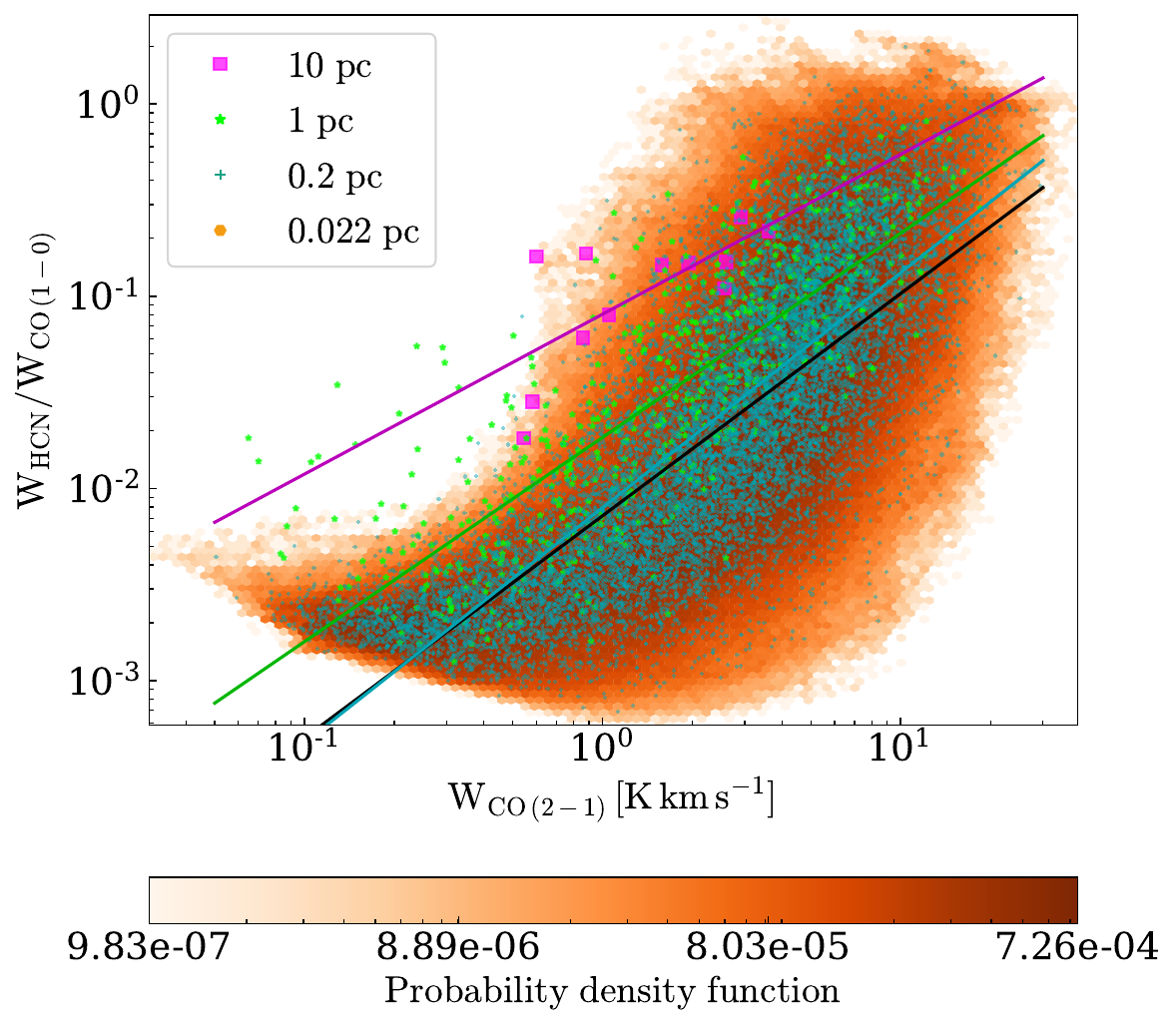}
    \caption{A plot of HCN/CO integrated intensity ratio against the velocity-integrated intensity of CO (2-1) at four different resolutions of 0.022, 0.2, 1 and 10 pc. Note that a cut-off of 0.001 $\mathrm{K\, km\, s^{-1}}$ is placed to simulate an observational cut-off due to noise.}    
    \label{fig:wcohcn}
\end{figure}

\begin{table}
	\centering
	\caption{Summary of our results of comparison between HCN/CO (1-0) to $\mathrm{W_{CO\, (2-1)}}$ for the 0.001 $\mathrm{K\, km\, s^{-1}}$. Note that these gradients are obtained through ordinary least squares (OLS) fitting.}
	\label{tab:tablehcnco}
	\begin{tabular}{lccr} 
		\hline
		& Resolution & Gradient of fit & \\
		& [$\mathrm{pc}$] & OLS & \\
		\hline
		& 0.022 & 1.16 & \\
		& 0.2 & 1.22 & \\
		& 1 & 1.06 & \\
		& 10 & 0.83 & \\
		\hline
	\end{tabular}
\end{table}

\begin{table}
	\centering
	\caption{Summary of our results of comparison between HCN/CO (1-0) to $\mathrm{W_{CO\, (2-1)}}$ for the 0.1 $\mathrm{K\, km\, s^{-1}}$. Note that these gradients are obtained through ordinary least squares (OLS) and total least squares (TLS) fitting.}
	\label{tab:tablehcncomulticol}
	\begin{tabular}{lcccr} 
		\hline
		& Resolution & \multicolumn{2}{c}{Gradient of fit} & \\
		& [$\mathrm{pc}$] & OLS & TLS & \\
		\hline
		& 0.022 & 0.52 & 5.1 & \\
		& 0.2 & 0.71 & 3.7 & \\
		& 1 & 0.67 & 1.83 & \\
		& 10 & 0.21 & 0.28 & \\
		\hline
	\end{tabular}
\end{table}

\begin{figure*}
    \centerline{
	    \includegraphics[width=80mm]{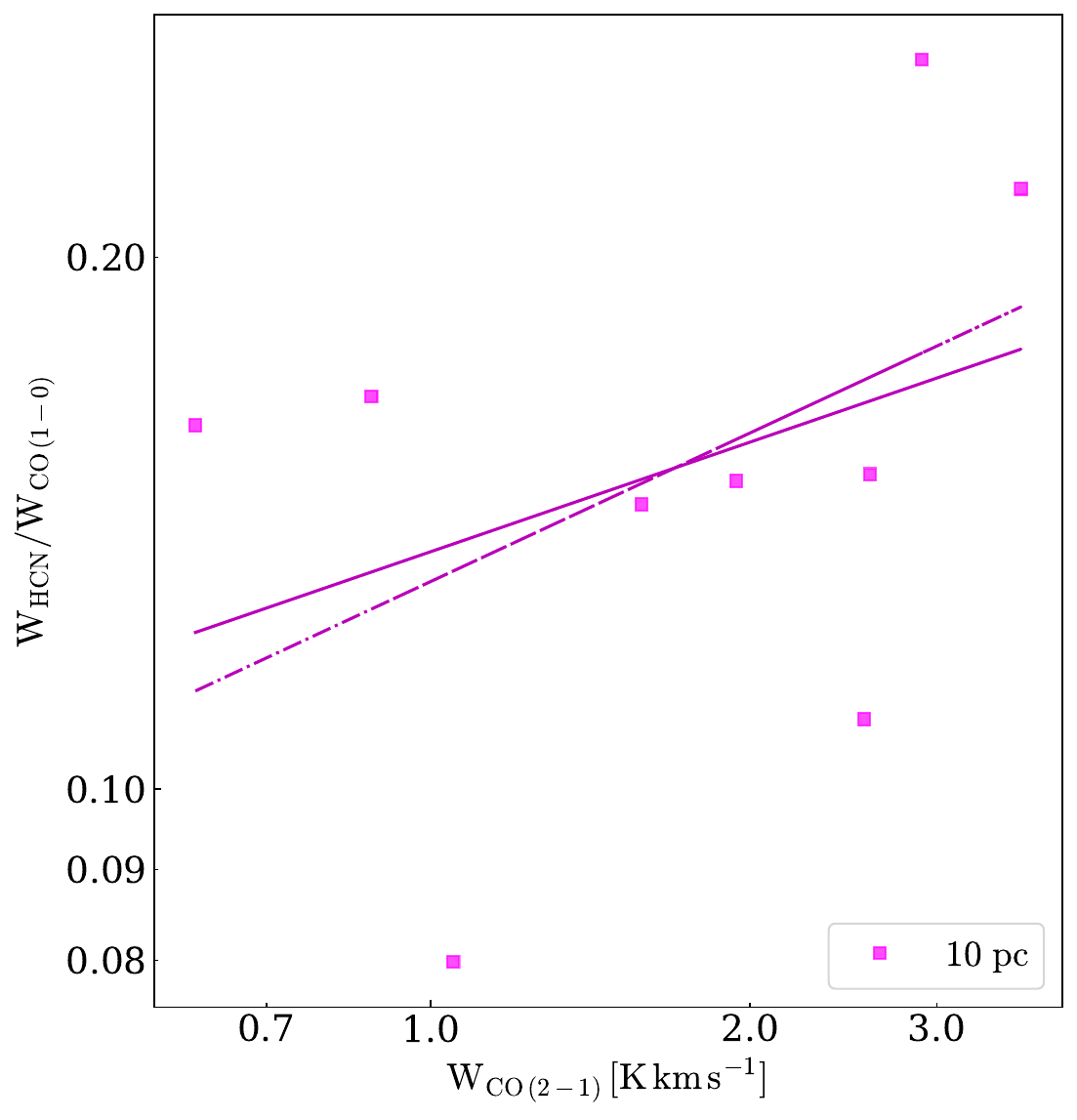}
	    \includegraphics[width=80mm]{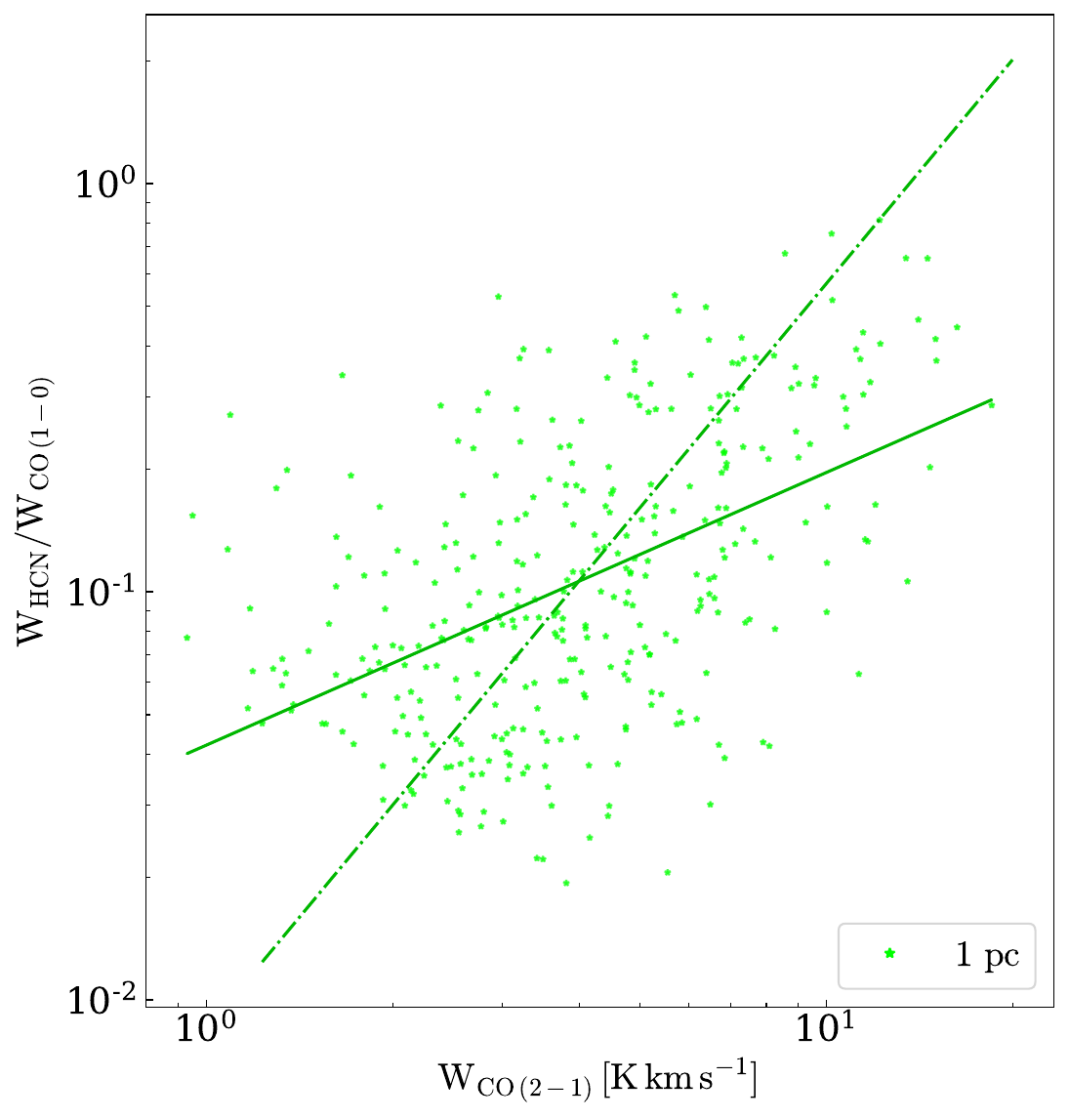}
    }
    \centerline{
	    \includegraphics[width=80mm]{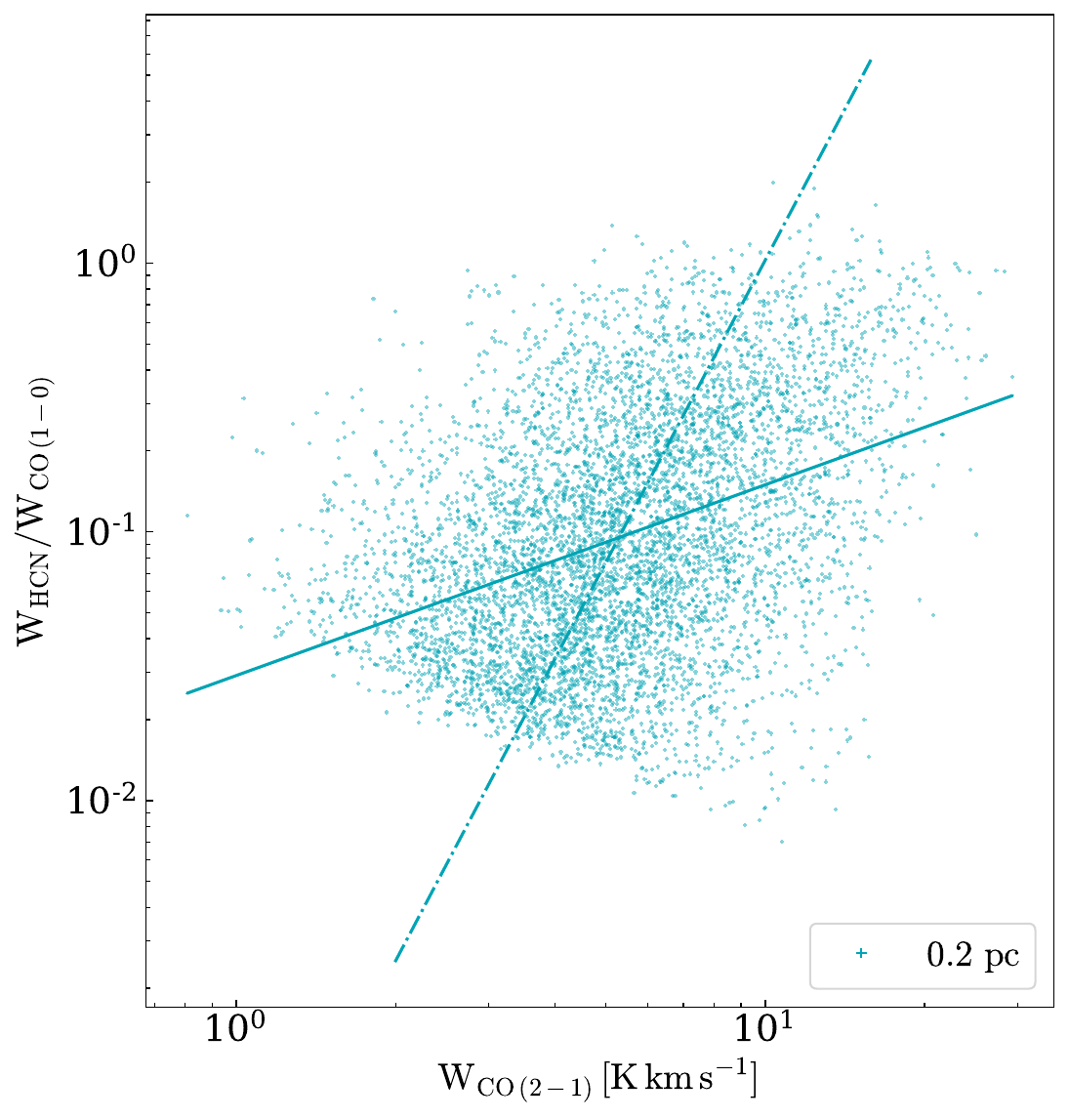}
	    \includegraphics[width=80mm]{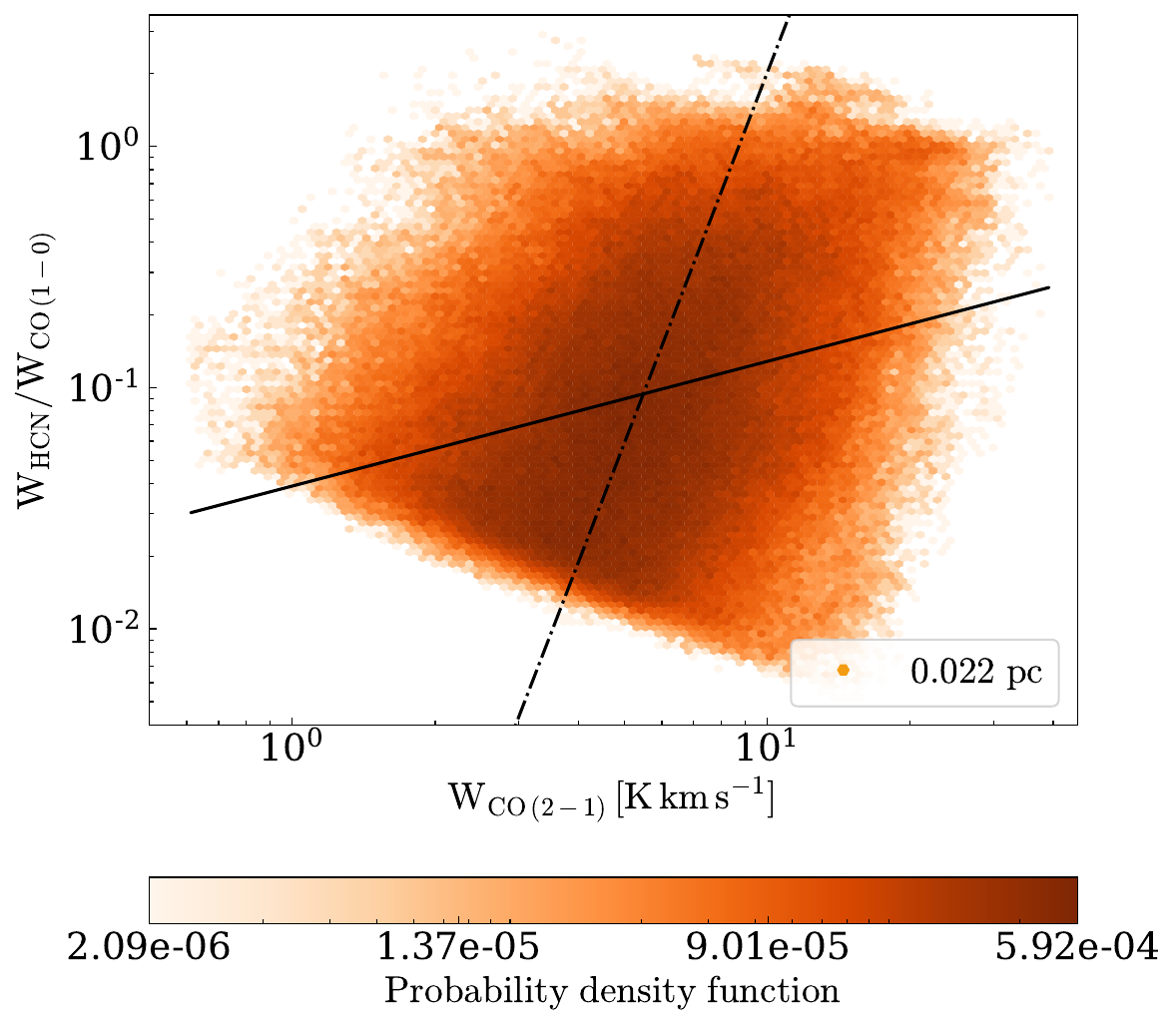}
    }
    \caption{As Figure \ref{fig:wcohcn}, but with a higher cut-off of 0.1 $\mathrm{K\, km\, s^{-1}}$. The dotted-dashed line denotes the gradient obtained through total least squares fitting and the solid line denotes the gradient obtained through ordinary least squares fitting. }
    \label{fig:wcohcnmulticol}
\end{figure*}

For this section we create twelve PPV cubes for all three lines (HCN (1-0), CO (1-0) and CO (2-1)) and for all twelve cases characterised by Table \ref{tab:tableoverview}. We create three additional resolutions for our analysis by degrading all twelve PPV cubes to 0.2, 1 and 10 pc respectively. The resulting variation in the resolution can be seen in Figure \ref{fig:NH2W}. A cut-off of 0.001 $\mathrm{K\, km\, s^{-1}}$ is placed on the velocity-integrated intensity of all three lines to limit our analysis to lines of sight where the abundances are more certain.

We present our HCN/CO integrated intensity ratio against $\mathrm{W_{CO\, (2-1)}}$ in Figure \ref{fig:wcohcn}, which despite the scatter, shows a clear correlation between $\mathrm{ W_{\rm HCN}/ W_{\rm CO}}$ and $\mathrm{W_{CO\, (2-1)}}$. We see that the relation between these observables becomes progressively more linear as the resolution decreases. To compare with \cite{GALLAGHER2018}, we present here a the results from a linear regression fit to the data in Figure \ref{fig:wcohcn}, assuming a power-law relationship between $\mathrm{ W_{\rm HCN}/ W_{\rm CO}}$ and $\mathrm{W_{CO\, (2-1)}}$; the gradient of fit for all four resolutions is given in Table \ref{tab:tablehcnco}. We see that as the resolution decreases (i.e. the pixel area increases) the gradient of fit also decreases (see Table \ref{tab:tablehcnco}). Although the trend is consistent with the \citet{GALLAGHER2018} study, who find a similar relationship between these quantities with a gradient of between 0.55 to 0.81 for resolutions in the range 650 - 2770 pc, the fact that the slope changes so much with resolution for the same underlying data,  implies there is no real physical justification for a linear relationship between these variables.  At very high resolution, that is, when the beam (each pixel) is fully filled with emission, we find a much steeper relationship than that seen in the unresolved, galactic-scale observations. Once again, this suggests that one needs to be extremely careful when interpreting data of this sort from galactic-scale surveys.

Our cut-off point of 0.001 $\mathrm{K\, km\, s^{-1}}$ in the emission used to generate Figure \ref{fig:wcohcn} is significantly lower than the $\sim$ 0.1 $\mathrm{K\, km\, s^{-1}}$  noise limit of observational surveys.  We therefore repeat the analysis above but with a higher velocity-integrated intensity cut-off of 0.1 $\mathrm{K\, km\, s^{-1}}$ on all three lines. For clarity, we plot the four different resolutions on four separate plots (see Figure \ref{fig:wcohcnmulticol}). With the higher cut-off, we lose the tail at low $\mathrm{W_{CO\, (2-1)}}$ that clearly has a very shallow gradient. We therefore would reasonably expect a steeper gradient for the higher cut-off analysis. However, using an ordinary least squares (OLS) fitting method we get much shallower gradients than our lower cut-off of 0.001 $\mathrm{K\, km\, s^{-1}}$ (see Table \ref{tab:tablehcncomulticol}), since it is sensitive to the outliers from the main trend (the main ridge that you can see by eye).  One could also argue that, observationally, there should be errors in both variables, and so OLS is not a statistically valid method for exploring a relation in this data.  We therefore compare the OLS to a total least squares (TLS) fitting method that minimizes the orthogonal difference of both the dependent and independent variables to the predicted best fit model. Using TLS to fit our data, we recover a best fit gradient that follows a high density ridge in the scatter. We now see much steeper gradients than what we saw for both OLS fitting of the cut-off of 0.001 $\mathrm{K\, km\, s^{-1}}$ and also 0.1 $\mathrm{K\, km\, s^{-1}}$.  We conclude here that the correlation in HCN/CO integrated intensity ratio against $\mathrm{W_{CO\, (2-1)}}$ in previous studies might be in fact due to the resolution and not necessarily due to a physical correlation. Indeed, our RT results would suggest that the underlying correlation between $\mathrm{ W_{HCN}/ W_{CO}}$ and $\mathrm{W_{CO\, (2-1)}}$ is significantly more complex.


%
%
%
\section{Discussion}
\label{sec:disc}

Although our work, and those of recent observational and numerical studies (\citealt{KAUFFMANN2017}, \citealt{PETY2017}, \citealt{ONUS2018}, \citealt{BARNES2020} and \citealt{EVANS2020}), suggests that HCN primarily traces lower density gas then previously assumed \citep{GAO2004a, KRUMHOLZ2007}, there is an undeniable correlation between HCN emission and star formation \citep{GAO2004a, GAO2004b}. More importantly, the correlation appears to be more linear than that between CO emission and star formation (e.g. \citealt{GAO2004b}). \citet{HACAR2020} has suggested that this correlation between HCN emission and star formation is due to the temperature dependence of the HCN to HNC abundance ratio, and that the ratio increases with temperature. In regions of star formation, where the interstellar radiation field is higher, we expect the gas to be hotter at the lower densities within the cloud (e.g. see \citealt{CLARK2019}). The combination of higher HCN abundance and higher temperature is then proposed to boost the emission. 

Although we do not include an explicitly temperature-dependent abundance in this paper, we can predict what the qualitative effect on our results would be.  If our clouds were exposed to a higher degree of ambient star formation, the gas at low densities would be hotter, and the HCN abundance would then be higher than what we currently adopt in these regions \citep{HACAR2020}. This would boost the emission from the lower density gas. The effect on our results would be, if anything, to lower the characteristic density traced by HCN emission, and thus this does not alter our main conclusion. However, whether this combination of effects leads to a linear relation between HCN emission and star formation remains to be tested, and we aim to revisit it in a future study.

Our use of the \citet{FUENTE2019} results to derive our HCN to CO abundance ratio also differs slightly from those of \citet{TAFALLA2021}. \citet{FUENTE2019} only observed a particular region of TMC-1 whilst \citet{TAFALLA2021} observed the entire cloud. However, using \citet{FUENTE2019} gives us a very conservative estimate for the HCN abundance at lower densities (below an $\mathrm{A_V}$ of 20 mag) compared to \citet{TAFALLA2021}. Once again, we can predict that using the  \citet{TAFALLA2021} abundance ration would almost certainly lower the density threshold ($\mathrm{n_{char}}$) as the higher HCN abundances at lower densities / $\mathrm{A_V}$ would provide a greater contribution of HCN emission to our Figure \ref{fig:fracemall} at lower densities.

The findings in this paper, and those from the observational studies of  \cite{PETY2017}, \cite{KAUFFMANN2017}, \cite{BARNES2020} and \cite{TAFALLA2021}, have serious implications for the use of HCN as a tracer of dense gas. This lower density threshold for HCN emission also has implications for the study of the star formation efficiency per free-fall time in GMCs. For example,  \cite{KRUMHOLZ2007} argue that star formation in dense gas is "slow", in the sense that only a small percentage of the gas forms stars every free-fall time. However, they assume that HCN emission traces gas with densities $\sim\, 6 \times 10^4\, \mathrm{cm^{-3}}$, which therefore has a short free-fall time. If the true value for the density traced by HCN is closer to the value of $\sim\, 3 \times 10^3\, \mathrm{cm^{-3}}$ that we find, this implies that the actual free-fall time is a factor of 4--5 longer than the value they derive, with a corresponding increase in the inferred star formation efficiency per free-fall time. Further, as we demonstrate at the end of Section \ref{subsubsec:HCNalpha}, the commonly-used methods for converting HCN emission into dense gas can over-predict the amount of gas residing at densities of $\rm n_{char}$ and higher be orders of magnitude, especially at early times when there is little star formation. This would again artificially lower the apparent star formation rate per free-fall time.



\section{Conclusions}
\label{sec:conc}

We investigate the relationship between gas density and HCN emission through post-processing of high resolution magnetohydrodynamical simulations of cloud-cloud collisions using RADMC-3D and {\sc Arepo}. We carry out 4 simulations with increasing cloud velocities from 1.875 $\mathrm{kms^{-1}}$ to 15 $\mathrm{kms^{-1}}$ and study the HCN emission from the clouds at 3 different times in each simulation, allowing us to explore a wide range of potential molecular cloud environments.

In our study, we find that HCN (1-0) emission traces gas with a characteristic volumetric density of $\mathrm{\sim 3 \times 10^3\ cm^{-3}}$ and a characteristic visual extinction of $\mathrm{\sim 5\ mag}$. Our findings are broadly consistent with those from recent observational studies \cite{PETY2017}, \cite{KAUFFMANN2017} and \cite{BARNES2020}, and taken together, implies that HCN emission traces more diffuse gas than previously thought (e.g. \citealt{GAO2004a}).  

We also find a luminosity to mass conversion factor of $\mathrm{ \alpha_{HCN}^{A_V}\ =\ 6.79\ M_{\odot}\, (K\, km\, s^{-1}\, pc^2)^{-1}}$ for $\mathrm{A_V}$ > 8 mag and $\mathrm{\alpha_{HCN}^{n_{\rm thr}} \ =\ 8.62\ M_{\odot}\, (K\, km\, s^{-1}\, pc^2)^{-1}}$ for $\mathrm{n\ >\ 2.85 \times 10^3\ cm^{-3}}$.  When we adopt the ``standard'' conversion factor with characteristic density $\mathrm{n\ >\ 3 \times 10^4\ cm^{-3}}$, we find that the analysis over-predicts the amount of ``dense'' gas by at least an order of magnitude. Indeed, in some cases, the conversion factor predicts gas in the the density range $\mathrm{n\ >\ 3 \times 10^4\ cm^{-3}}$ when no gas above that density exists in our simulations. 

\section*{Acknowledgements}

We thank the anonymous reviewer for constructive comments that greatly improved the paper. This research was undertaken using the supercomputing facilities at Cardiff University operated by Advanced Research Computing at Cardiff (ARCCA) on behalf of the Cardiff Supercomputing Facility and the HPC Wales and Supercomputing Wales (SCW) projects. We acknowledge the support of the latter, which is part-funded by the European Regional Development Fund (ERDF) via the Welsh Government’.  PCC gratefully acknowledges the support of a STFC Consolidated Grant (ST/K00926/1), and support from the StarFormMapper project, funded by the European Union's Horizon 2020 research and innovation programme under grant agreement No 687528. SCOG acknowledges support from the DFG via SFB 881 ``The Milky Way System'' (sub-projects B1, B2 and B8), from the Heidelberg cluster of excellence EXC 2181-390900948 ``STRUCTURES: A unifying approach to emergent phenomena in the physical world, mathematics, and complex data'', funded by the German Excellence Strategy, and from the European Research Council via the ERC Synergy Grant 1798 ``ECOGAL – Understanding our Galactic ecosystem: From the disk of the Milky Way to the formation sites of stars and planets'' (project ID 855130). This project has received funding from the European Research Council (ERC) under the European Union’s Horizon 2020 research and innovation programme (Grant agreement Nos. 851435 and 639459).



\section*{Data Availability}

The data underlying this article will be shared on reasonable request
to the corresponding author.



\bibliographystyle{mnras}
\bibliography{HCN_Jones.bib}

\begin{thebibliography}{}
\makeatletter
\relax
\def\mn@urlcharsother{\let\do\@makeother \do\$\do\&\do\#\do\^\do\_\do\%\do\~}
\def\mn@doi{\begingroup\mn@urlcharsother \@ifnextchar [ {\mn@doi@}
  {\mn@doi@[]}}
\def\mn@doi@[#1]#2{\def\@tempa{#1}\ifx\@tempa\@empty \href
  {http://dx.doi.org/#2} {doi:#2}\else \href {http://dx.doi.org/#2} {#1}\fi
  \endgroup}
\def\mn@eprint#1#2{\mn@eprint@#1:#2::\@nil}
\def\mn@eprint@arXiv#1{\href {http://arxiv.org/abs/#1} {{\tt arXiv:#1}}}
\def\mn@eprint@dblp#1{\href {http://dblp.uni-trier.de/rec/bibtex/#1.xml}
  {dblp:#1}}
\def\mn@eprint@#1:#2:#3:#4\@nil{\def\@tempa {#1}\def\@tempb {#2}\def\@tempc
  {#3}\ifx \@tempc \@empty \let \@tempc \@tempb \let \@tempb \@tempa \fi \ifx
  \@tempb \@empty \def\@tempb {arXiv}\fi \@ifundefined
  {mn@eprint@\@tempb}{\@tempb:\@tempc}{\expandafter \expandafter \csname
  mn@eprint@\@tempb\endcsname \expandafter{\@tempc}}}

\bibitem[\protect\citeauthoryear{Barnes et~al.,}{Barnes
  et~al.}{2020}]{BARNES2020}
Barnes A.~T.,  et~al., 2020, \mn@doi [Monthly Notices of the Royal Astronomical
  Society] {10.1093/mnras/staa1814}, 497, 1972

\bibitem[\protect\citeauthoryear{Bate, Bonnell  \& Price}{Bate
  et~al.}{1995}]{BATE1995}
Bate M.~R.,  Bonnell I.~A.,   Price N.~M.,  1995, \mn@doi [Monthly Notices of
  the Royal Astronomical Society] {10.1093/mnras/277.2.362}, 277, 362

\bibitem[\protect\citeauthoryear{Bigiel, Walter, Blitz, Brinks, de Blok  \&
  Madore}{Bigiel et~al.}{2010}]{BIGIEL2010}
Bigiel F.,  Walter F.,  Blitz L.,  Brinks E.,  de Blok W. J.~G.,   Madore B.,
  2010, \mn@doi [The Astronomical Journal] {10.1088/0004-6256/140/5/1194}, 140,
  1194

\bibitem[\protect\citeauthoryear{Bohlin, Savage  \& Drake}{Bohlin
  et~al.}{1978}]{BOHLIN1978}
Bohlin R.~C.,  Savage B.~D.,   Drake J.~F.,  1978, \mn@doi [The Astrophysical
  Journal] {10.1086/156357}, 224, 132

\bibitem[\protect\citeauthoryear{Clark, Glover  \& Klessen}{Clark
  et~al.}{2012}]{CLARK2012TREECOL}
Clark P.~C.,  Glover S. C.~O.,   Klessen R.~S.,  2012, \mn@doi [Monthly Notices
  of the Royal Astronomical Society] {10.1111/j.1365-2966.2011.20087.x}, 420,
  745

\bibitem[\protect\citeauthoryear{Clark, Glover, Ragan  \&
  {Duarte-Cabral}}{Clark et~al.}{2019}]{CLARK2019}
Clark P.~C.,  Glover S. C.~O.,  Ragan S.~E.,   {Duarte-Cabral} A.,  2019,
  \mn@doi [Monthly Notices of the Royal Astronomical Society]
  {10.1093/mnras/stz1119}, 486, 4622

\bibitem[\protect\citeauthoryear{Crutcher}{Crutcher}{2012}]{CRUTCHER2012}
Crutcher R.~M.,  2012, \mn@doi [Annual Review of Astronomy and Astrophysics]
  {10.1146/annurev-astro-081811-125514}, 50, 29

\bibitem[\protect\citeauthoryear{Crutcher, Wandelt, Heiles, Falgarone  \&
  Troland}{Crutcher et~al.}{2010}]{CRUTCHER2010}
Crutcher R.~M.,  Wandelt B.,  Heiles C.,  Falgarone E.,   Troland T.~H.,  2010,
  \mn@doi [The Astrophysical Journal] {10.1088/0004-637X/725/1/466}, 725, 466

\bibitem[\protect\citeauthoryear{Dedner, Kemm, Kr{\"o}ner, Munz, Schnitzer  \&
  Wesenberg}{Dedner et~al.}{2002}]{DEDNER2002}
Dedner A.,  Kemm F.,  Kr{\"o}ner D.,  Munz C.~D.,  Schnitzer T.,   Wesenberg
  M.,  2002, \mn@doi [Journal of Computational Physics]
  {10.1006/jcph.2001.6961}, 175, 645

\bibitem[\protect\citeauthoryear{Dobbs}{Dobbs}{2008}]{DOBBS2008}
Dobbs C.~L.,  2008, \mn@doi [Monthly Notices of the Royal Astronomical Society]
  {10.1111/j.1365-2966.2008.13939.x}, 391, 844

\bibitem[\protect\citeauthoryear{Draine}{Draine}{1978}]{DRAINE1978}
Draine B.~T.,  1978, \mn@doi [The Astrophysical Journal Supplement Series]
  {10.1086/190513}, 36, 595

\bibitem[\protect\citeauthoryear{Draine}{Draine}{2011}]{DRAINE2011}
Draine B.~T.,  2011, Physics of the Interstellar and Intergalactic Medium by
  Bruce T. Draine. Princeton University Press, 2011. ISBN: 978-0-691-12214-4

\bibitem[\protect\citeauthoryear{Draine \& Bertoldi}{Draine \&
  Bertoldi}{1996}]{DRAINE1996}
Draine B.~T.,  Bertoldi F.,  1996, \mn@doi [The Astrophysical Journal]
  {10.1086/177689}, 468, 269

\bibitem[\protect\citeauthoryear{Dullemond, Juhasz, Pohl, Sereshti, Shetty,
  Peters, Commercon  \& Flock}{Dullemond et~al.}{2012}]{DULLEMOND2012}
Dullemond C.~P.,  Juhasz A.,  Pohl A.,  Sereshti F.,  Shetty R.,  Peters T.,
  Commercon B.,   Flock M.,  2012, Astrophysics Source Code Library, p.
  ascl:1202.015

\bibitem[\protect\citeauthoryear{Dumouchel, Faure  \& Lique}{Dumouchel
  et~al.}{2010}]{DUMOUCHEL2010}
Dumouchel F.,  Faure A.,   Lique F.,  2010, \mn@doi [Monthly Notices of the
  Royal Astronomical Society] {10.1111/j.1365-2966.2010.16826.x}, 406, 2488

\bibitem[\protect\citeauthoryear{Elmegreen}{Elmegreen}{1994}]{ELMEGREEN1994}
Elmegreen B.~G.,  1994, \mn@doi [The Astrophysical Journal Letters]
  {10.1086/187313}, 425, L73

\bibitem[\protect\citeauthoryear{Evans, Kim, Wu, Chao, Heyer, Liu,
  {Nguyen-Lu'o'ng}  \& Kauffmann}{Evans et~al.}{2020}]{EVANS2020}
Evans N.~J.,  Kim K.-T.,  Wu J.,  Chao Z.,  Heyer M.,  Liu T.,
  {Nguyen-Lu'o'ng} Q.,   Kauffmann J.,  2020, \mn@doi [The Astrophysical
  Journal] {10.3847/1538-4357/ab8938}, 894, 103

\bibitem[\protect\citeauthoryear{Faure, Varambhia, Stoecklin  \&
  Tennyson}{Faure et~al.}{2007}]{FAURE2007}
Faure A.,  Varambhia H.~N.,  Stoecklin T.,   Tennyson J.,  2007, \mn@doi
  [Monthly Notices of the Royal Astronomical Society]
  {10.1111/j.1365-2966.2007.12416.x}, 382, 840

\bibitem[\protect\citeauthoryear{Federrath}{Federrath}{2015}]{FEDERRATH2015}
Federrath C.,  2015, \mn@doi [Monthly Notices of the Royal Astronomical
  Society] {10.1093/mnras/stv941}, 450, 4035

\bibitem[\protect\citeauthoryear{Federrath, Banerjee, Clark  \&
  Klessen}{Federrath et~al.}{2010}]{FEDERRATH2010}
Federrath C.,  Banerjee R.,  Clark P.~C.,   Klessen R.~S.,  2010, \mn@doi [The
  Astrophysical Journal] {10.1088/0004-637X/713/1/269}, 713, 269

\bibitem[\protect\citeauthoryear{Fuente et~al.,}{Fuente
  et~al.}{2019}]{FUENTE2019}
Fuente A.,  et~al., 2019, \mn@doi [Astronomy \& Astrophysics]
  {10.1051/0004-6361/201834654}, 624, A105

\bibitem[\protect\citeauthoryear{Gallagher et~al.,}{Gallagher
  et~al.}{2018}]{GALLAGHER2018}
Gallagher M.~J.,  et~al., 2018, \mn@doi [The Astrophysical Journal]
  {10.3847/2041-8213/aaf16a}, 868, L38

\bibitem[\protect\citeauthoryear{Gao \& Solomon}{Gao \&
  Solomon}{2004a}]{GAO2004a}
Gao Y.,  Solomon P.~M.,  2004a, \mn@doi [The Astrophysical Journal Supplement
  Series] {10.1086/383003}, 152, 63

\bibitem[\protect\citeauthoryear{Gao \& Solomon}{Gao \&
  Solomon}{2004b}]{GAO2004b}
Gao Y.,  Solomon P.~M.,  2004b, \mn@doi [The Astrophysical Journal]
  {10.1086/382999}, 606, 271

\bibitem[\protect\citeauthoryear{Gao, Carilli, Solomon  \& Bout}{Gao
  et~al.}{2007}]{GAO2007}
Gao Y.,  Carilli C.~L.,  Solomon P.~M.,   Bout P. A.~V.,  2007, \mn@doi [The
  Astrophysical Journal] {10.1086/518244}, 660, L93

\bibitem[\protect\citeauthoryear{{Garc{\'i}a-Burillo}, Usero, {Alonso-Herrero},
  {Graci{\'a}-Carpio}, {Pereira-Santaella}, Colina, Planesas  \&
  Arribas}{{Garc{\'i}a-Burillo} et~al.}{2012}]{GARCIA-BURILLO2012}
{Garc{\'i}a-Burillo} S.,  Usero A.,  {Alonso-Herrero} A.,  {Graci{\'a}-Carpio}
  J.,  {Pereira-Santaella} M.,  Colina L.,  Planesas P.,   Arribas S.,  2012,
  \mn@doi [Astronomy \& Astrophysics] {10.1051/0004-6361/201117838}, 539, A8

\bibitem[\protect\citeauthoryear{Glover \& Clark}{Glover \&
  Clark}{2012}]{GLOVER2012}
Glover S. C.~O.,  Clark P.~C.,  2012, \mn@doi [Monthly Notices of the Royal
  Astronomical Society] {10.1111/j.1365-2966.2011.19648.x}, 421, 9

\bibitem[\protect\citeauthoryear{Glover \& Mac~Low}{Glover \&
  Mac~Low}{2007}]{GLOVER2007}
Glover S. C.~O.,  Mac~Low M.-M.,  2007, \mn@doi [The Astrophysical Journal]
  {10.1086/512227}, 659, 1317

\bibitem[\protect\citeauthoryear{Habing}{Habing}{1968}]{HABING1968}
Habing H.~J.,  1968, Bulletin of the Astronomical Institutes of the
  Netherlands, 19, 421

\bibitem[\protect\citeauthoryear{Hacar, Bosman  \& {van Dishoeck}}{Hacar
  et~al.}{2020}]{HACAR2020}
Hacar A.,  Bosman A.~D.,   {van Dishoeck} E.~F.,  2020, \mn@doi [Astronomy \&
  Astrophysics] {10.1051/0004-6361/201936516}, 635, A4

\bibitem[\protect\citeauthoryear{Harada, Nishimura, Watanabe, Yamamoto, Aikawa,
  Sakai  \& Shimonishi}{Harada et~al.}{2019}]{HARADA2019}
Harada N.,  Nishimura Y.,  Watanabe Y.,  Yamamoto S.,  Aikawa Y.,  Sakai N.,
  Shimonishi T.,  2019, \mn@doi [The Astrophysical Journal]
  {10.3847/1538-4357/aaf72a}, 871, 238

\bibitem[\protect\citeauthoryear{Hunter, Clark, Glover  \& Klessen}{Hunter
  et~al.}{2021}]{HUNTER2021}
Hunter G.~H.,  Clark P.~C.,  Glover S. C.~O.,   Klessen R.~S.,  2021,
  arXiv:2109.06195 [astro-ph]

\bibitem[\protect\citeauthoryear{Jankowski \& Szalewicz}{Jankowski \&
  Szalewicz}{2005}]{JANKOWSKI2005}
Jankowski P.,  Szalewicz K.,  2005, \mn@doi [Journal of Chemical Physics]
  {10.1063/1.2008216}, 123, 104301

\bibitem[\protect\citeauthoryear{Kainulainen, Beuther, Henning  \&
  Plume}{Kainulainen et~al.}{2009}]{KAINULAINEN2009}
Kainulainen J.,  Beuther H.,  Henning T.,   Plume R.,  2009, \mn@doi [Astronomy
  and Astrophysics] {10.1051/0004-6361/200913605}, 508, L35

\bibitem[\protect\citeauthoryear{Kauffmann, Goldsmith, Melnick, Tolls, Guzman
  \& Menten}{Kauffmann et~al.}{2017}]{KAUFFMANN2017}
Kauffmann J.,  Goldsmith P.~F.,  Melnick G.,  Tolls V.,  Guzman A.,   Menten
  K.~M.,  2017, \mn@doi [Astronomy and Astrophysics]
  {10.1051/0004-6361/201731123}, 605, L5

\bibitem[\protect\citeauthoryear{Kennicutt}{Kennicutt}{1989}]{KENNICUTT1989}
Kennicutt Jr. R.~C.,  1989, \mn@doi [The Astrophysical Journal]
  {10.1086/167834}, 344, 685

\bibitem[\protect\citeauthoryear{Kennicutt \& Evans}{Kennicutt \&
  Evans}{2012}]{KENNICUTT2012}
Kennicutt R.~C.,  Evans N.~J.,  2012, \mn@doi [Annual Review of Astronomy and
  Astrophysics] {10.1146/annurev-astro-081811-125610}, 50, 531

\bibitem[\protect\citeauthoryear{Kruijssen, Longmore, Elmegreen, Murray, Bally,
  Testi  \& Kennicutt}{Kruijssen et~al.}{2014}]{KRUIJSSEN2014}
Kruijssen J. M.~D.,  Longmore S.~N.,  Elmegreen B.~G.,  Murray N.,  Bally J.,
  Testi L.,   Kennicutt Jr R.~C.,  2014, \mn@doi [Monthly Notices of the Royal
  Astronomical Society] {10.1093/mnras/stu494}, 440, 3370

\bibitem[\protect\citeauthoryear{Krumholz \& Tan}{Krumholz \&
  Tan}{2007}]{KRUMHOLZ2007}
Krumholz M.~R.,  Tan J.~C.,  2007, \mn@doi [The Astrophysical Journal]
  {10.1086/509101}, 654, 304

\bibitem[\protect\citeauthoryear{Krumholz \& Thompson}{Krumholz \&
  Thompson}{2007}]{KRUMHOLZ2007a}
Krumholz M.~R.,  Thompson T.~A.,  2007, \mn@doi [The Astrophysical Journal]
  {10.1086/521642}, 669, 289

\bibitem[\protect\citeauthoryear{Lada, Lombardi  \& Alves}{Lada
  et~al.}{2010}]{LADA2010}
Lada C.~J.,  Lombardi M.,   Alves J.~F.,  2010, \mn@doi [The Astrophysical
  Journal] {10.1088/0004-637X/724/1/687}, 724, 687

\bibitem[\protect\citeauthoryear{Lada, Forbrich, Lombardi  \& Alves}{Lada
  et~al.}{2012}]{LADA2012}
Lada C.~J.,  Forbrich J.,  Lombardi M.,   Alves J.~F.,  2012, \mn@doi [The
  Astrophysical Journal] {10.1088/0004-637X/745/2/190}, 745, 190

\bibitem[\protect\citeauthoryear{Leroy et~al.,}{Leroy et~al.}{2017}]{LEROY2017}
Leroy A.~K.,  et~al., 2017, \mn@doi [The Astrophysical Journal]
  {10.3847/1538-4357/835/2/217}, 835, 217

\bibitem[\protect\citeauthoryear{Longmore et~al.,}{Longmore
  et~al.}{2013}]{LONGMORE2013}
Longmore S.~N.,  et~al., 2013, \mn@doi [Monthly Notices of the Royal
  Astronomical Society] {10.1093/mnras/sts376}, 429, 987

\bibitem[\protect\citeauthoryear{Mathis, Mezger  \& Panagia}{Mathis
  et~al.}{1983}]{MATHIS1983}
Mathis J.~S.,  Mezger P.~G.,   Panagia N.,  1983, Astronomy and Astrophysics,
  128, 212

\bibitem[\protect\citeauthoryear{Nguyen, Jackson, Henkel, Truong  \&
  Mauersberger}{Nguyen et~al.}{1992}]{NGUYEN1992}
Nguyen Q.-R.,  Jackson J.~M.,  Henkel C.,  Truong B.,   Mauersberger R.,  1992,
  \mn@doi [The Astrophysical Journal] {10.1086/171944}, 399, 521

\bibitem[\protect\citeauthoryear{Onus, Krumholz  \& Federrath}{Onus
  et~al.}{2018}]{ONUS2018}
Onus A.,  Krumholz M.~R.,   Federrath C.,  2018, \mn@doi [Monthly Notices of
  the Royal Astronomical Society] {10.1093/mnras/sty1662}, 479, 1702

\bibitem[\protect\citeauthoryear{Pakmor, Bauer  \& Springel}{Pakmor
  et~al.}{2011}]{PAKMOR2011}
Pakmor R.,  Bauer A.,   Springel V.,  2011, \mn@doi [Monthly Notices of the
  Royal Astronomical Society] {10.1111/j.1365-2966.2011.19591.x}, 418, 1392

\bibitem[\protect\citeauthoryear{Pety et~al.,}{Pety et~al.}{2017}]{PETY2017}
Pety J.,  et~al., 2017, \mn@doi [Astronomy and Astrophysics]
  {10.1051/0004-6361/201629862}, 599, A98

\bibitem[\protect\citeauthoryear{Powell, Roe, Linde, Gombosi  \&
  De~Zeeuw}{Powell et~al.}{1999}]{POWELL1999}
Powell K.~G.,  Roe P.~L.,  Linde T.~J.,  Gombosi T.~I.,   De~Zeeuw D.~L.,
  1999, \mn@doi [Journal of Computational Physics] {10.1006/jcph.1999.6299},
  154, 284

\bibitem[\protect\citeauthoryear{Priestley \& Whitworth}{Priestley \&
  Whitworth}{2020}]{PRIESTLEY2020}
Priestley F.~D.,  Whitworth A.~P.,  2020, \mn@doi [Monthly Notices of the Royal
  Astronomical Society] {10.1093/mnras/staa3111}, 499, 3728

\bibitem[\protect\citeauthoryear{Priestley \& Whitworth}{Priestley \&
  Whitworth}{2021}]{PRIESTLEY2021}
Priestley F.~D.,  Whitworth A.~P.,  2021, \mn@doi [Monthly Notices of the Royal
  Astronomical Society] {10.1093/mnras/stab1777}, 506, 775

\bibitem[\protect\citeauthoryear{Querejeta et~al.,}{Querejeta
  et~al.}{2019}]{QUEREJETA2019}
Querejeta M.,  et~al., 2019, \mn@doi [Astronomy \& Astrophysics]
  {10.1051/0004-6361/201834915}, 625, A19

\bibitem[\protect\citeauthoryear{Riechers, Walter, Carilli, Weiss, Bertoldi,
  Menten, Knudsen  \& Cox}{Riechers et~al.}{2006}]{RIECHERS2006}
Riechers D.~A.,  Walter F.,  Carilli C.~L.,  Weiss A.,  Bertoldi F.,  Menten
  K.~M.,  Knudsen K.~K.,   Cox P.,  2006, \mn@doi [The Astrophysical Journal
  Letters] {10.1086/505908}, 645, L13

\bibitem[\protect\citeauthoryear{Schmidt}{Schmidt}{1959}]{SCHMIDT1959}
Schmidt M.,  1959, \mn@doi [The Astrophysical Journal] {10.1086/146614}, 129,
  243

\bibitem[\protect\citeauthoryear{Sch{\"o}ier, {van der Tak}, {van Dishoeck}  \&
  Black}{Sch{\"o}ier et~al.}{2005}]{SCHOIER2005}
Sch{\"o}ier F.~L.,  {van der Tak} F. F.~S.,  {van Dishoeck} E.~F.,   Black
  J.~H.,  2005, \mn@doi [Astronomy \& Astrophysics]
  {10.1051/0004-6361:20041729}, 432, 369

\bibitem[\protect\citeauthoryear{Sembach, Howk, Ryans  \& Keenan}{Sembach
  et~al.}{2000}]{SEMBACH2000}
Sembach K.~R.,  Howk J.~C.,  Ryans R. S.~I.,   Keenan F.~P.,  2000, \mn@doi
  [The Astrophysical Journal] {10.1086/308173}, 528, 310

\bibitem[\protect\citeauthoryear{Shetty, Glover, Dullemond  \& Klessen}{Shetty
  et~al.}{2011}]{SHETTY2011}
Shetty R.,  Glover S.~C.,  Dullemond C.~P.,   Klessen R.~S.,  2011, \mn@doi
  [Monthly Notices of the Royal Astronomical Society]
  {10.1111/j.1365-2966.2010.18005.x}, 412, 1686

\bibitem[\protect\citeauthoryear{Shetty, Kelly, Rahman, Bigiel, Bolatto, Clark,
  Klessen  \& Konstandin}{Shetty et~al.}{2014}]{SHETTY2014}
Shetty R.,  Kelly B.~C.,  Rahman N.,  Bigiel F.,  Bolatto A.~D.,  Clark P.~C.,
  Klessen R.~S.,   Konstandin L.~K.,  2014, \mn@doi [Monthly Notices of the
  Royal Astronomical Society] {10.1093/mnrasl/slt156}, 437, L61

\bibitem[\protect\citeauthoryear{Shirley}{Shirley}{2015}]{SHIRLEY2015}
Shirley Y.~L.,  2015, \mn@doi [Publications of the Astronomical Society of the
  Pacific] {10.1086/680342}, 127, 299

\bibitem[\protect\citeauthoryear{Sobolev}{Sobolev}{1957}]{SOBOLEV1957}
Sobolev V.~V.,  1957, Soviet Astronomy, 1, 678

\bibitem[\protect\citeauthoryear{Springel}{Springel}{2010}]{SPRINGEL2010}
Springel V.,  2010, \mn@doi [Monthly Notices of the Royal Astronomical Society]
  {10.1111/j.1365-2966.2009.15715.x}, 401, 791

\bibitem[\protect\citeauthoryear{Tafalla, Usero  \& Hacar}{Tafalla
  et~al.}{2021}]{TAFALLA2021}
Tafalla M.,  Usero A.,   Hacar A.,  2021, \mn@doi [Astronomy \& Astrophysics]
  {10.1051/0004-6361/202038727}, 646, A97

\bibitem[\protect\citeauthoryear{Tress, Smith, Sormani, Glover, Klessen,
  Mac~Low  \& Clark}{Tress et~al.}{2020}]{TRESS2020}
Tress R.~G.,  Smith R.~J.,  Sormani M.~C.,  Glover S. C.~O.,  Klessen R.~S.,
  Mac~Low M.-M.,   Clark P.~C.,  2020, \mn@doi [Monthly Notices of the Royal
  Astronomical Society] {10.1093/mnras/stz3600}, 492, 2973

\bibitem[\protect\citeauthoryear{Weinberger, Springel  \& Pakmor}{Weinberger
  et~al.}{2020}]{WEINBERGER2020}
Weinberger R.,  Springel V.,   Pakmor R.,  2020, \mn@doi [The Astrophysical
  Journal Supplement Series] {10.3847/1538-4365/ab908c}, 248, 32

\bibitem[\protect\citeauthoryear{Wollenberg, Glover, Clark  \&
  Klessen}{Wollenberg et~al.}{2020}]{WOLLENBERG2020}
Wollenberg K. M.~J.,  Glover S. C.~O.,  Clark P.~C.,   Klessen R.~S.,  2020,
  \mn@doi [Monthly Notices of the Royal Astronomical Society]
  {10.1093/mnras/staa289}, 494, 1871

\bibitem[\protect\citeauthoryear{Wong \& Blitz}{Wong \& Blitz}{2002}]{WONG2002}
Wong T.,  Blitz L.,  2002, \mn@doi [The Astrophysical Journal]
  {10.1086/339287}, 569, 157

\bibitem[\protect\citeauthoryear{Wu, Evans, Gao, Solomon, Shirley  \&
  Vanden~Bout}{Wu et~al.}{2005}]{WU2005}
Wu J.,  Evans II N.~J.,  Gao Y.,  Solomon P.~M.,  Shirley Y.~L.,   Vanden~Bout
  P.~A.,  2005, \mn@doi [The Astrophysical Journal Letters] {10.1086/499623},
  635, L173

\bibitem[\protect\citeauthoryear{Yang, Stancil, Balakrishnan  \& Forrey}{Yang
  et~al.}{2010}]{YANG2010}
Yang B.,  Stancil P.~C.,  Balakrishnan N.,   Forrey R.~C.,  2010, \mn@doi [The
  Astrophysical Journal] {10.1088/0004-637X/718/2/1062}, 718, 1062

\makeatother
\end{thebibliography}




\bsp	
\label{lastpage}
\end{document}